\documentclass[12pt]{article}
\usepackage{epsfig}
\usepackage{amsmath}
\usepackage{amssymb}
\usepackage{axodraw}
\usepackage{booktabs}

\def\lsim{\mathrel{\rlap{\lower4pt\hbox{\hskip1pt$\sim$}}
    \raise1pt\hbox{$<$}}}                
\def\gsim{\mathrel{\rlap{\lower4pt\hbox{\hskip1pt$\sim$}}
    \raise1pt\hbox{$>$}}}                

\newcommand{\alphas}{\alpha_{\mathrm{s}}}

\newcommand{\p}{\mathrm{p}}
\newcommand{\q}{\mathrm{q}}

\newcommand{\W}{\mathrm{W}}

{\end{list}}
\newcounter{enumct}

\newlength{\abstwidth}
\newlength{\captivewidth}
\begin{document}
 
\sloppy

\pagestyle{empty}

\begin{flushright}
LU TP 03--03\\
hep-ph/0302225\\
February 2003
\end{flushright}

\vspace{\fill}

\begin{center}
{\LARGE\bf A Position-Dependent}\\[3mm]
{\LARGE\bf Bose--Einstein Correlation Model}\\[5mm]
{\LARGE\bf for Hadronization Processes}\\[10mm]
{\Large Thesis for Master of Science Degree}\\[3mm]
{\Large Henrik Johansson\footnote{henrikj@thep.lu.se}}\\[3mm]
{\it Department of Theoretical Physics,}\\[1mm]
{\it Lund University,}\\[1mm]
{\it S\"olvegatan 14A,}\\[1mm]
{\it S-223 62 Lund, Sweden}
\end{center}
 
\vspace{\fill}
 
\begin{center}
{\bf Abstract}\\[2ex]
\begin{minipage}{\abstwidth}
Particles are either bosons or fermions, which obey Bose--Einstein (BE) or Fermi--Dirac statistics, respectively. In the hadronization process the effects of the former kind are the most important ones, since mesons are the most common products in high-energy collisions. In the phenomenological Lund Model the BE effect is approximated by a semiclassical momentum dependent correlation function, which effectively acts as an attractive force between two mesons. 
However, the Lund Model does also provide a space--time picture of the hadronization process. Based on this, we have developed a simple extension of the current LUBOEI algorithm that makes use of the production vertex distance between two mesons. 

Comparisons between the old model and the new one is made for $\pi$-mesons according to exponential and Gaussian parameterizations of the BE correlation function. Qualitatively we note a relative suppression of the BE effects in multi-jet events, and especially for pairs of low momenta, with the new algorithm. This trend is in agreement with preliminary experimental data. 
\end{minipage}
\end{center}
 
\vspace{\fill}

\clearpage
\pagestyle{plain}
\setcounter{page}{1} 

\section{Introduction}

Like the Greeks thought of the world as consisting of indivisible parts called atoms, we today associate matter and forces with quantized objects simply called \emph{particles}; or, in the mathematical language of Quantum Field Theory (QFT), particles are associated with excitations of \emph{fields}. The theory that describes the subatomic world, as we know it today, is called the \emph{Standard Model}. The simplicity of this model comes from the fact that there are only a handful of fundamental particles, such as \emph{quarks} and \emph{leptons}, which make up matter,  and \emph{gauge bosons}, which provide the forces or interactions. The complexity of our world comes from the fact that particles can take on different properties called quantum numbers, but more importantly they can be combined into larger units according to the rules set by the interactions. Combining three quarks one obtains \emph{baryons}, combining a quark and an antiquark one obtains \emph{mesons}. Mesons and baryons are members of a larger class of particles called \emph{hadrons}. The combination of quarks into more complex particles cannot be made arbitrary, since they are intimately controlled by the interactions; we will come back to this later. Of course, the everyday particles, that we are more familiar with, are not hadrons or leptons, but atoms that consists of protons, neutrons (baryons) and electrons (leptons), and molecules that are made up by several atoms. This last subject, however interesting it may be, is outside the scope of this thesis.

Out of the four fundamental interactions only three are included in the Standard Model; the fourth, \emph{gravitation}, still refuses to give in to all attempts of quantization that are based on standard gauge theories. However important in the everyday world, it is weaker than all the other forces in the microscopic world and can be neglected in the following. \emph{Electromagnetism} and the \emph{strong force} are treated in the theories QED (Quantum Electro-Dynamics) and QCD (Quantum Chromo-Dynamics) respectively. The \emph{weak force} has no separate theory and was understood only when it was unified with electromagnetism, in what is known as the electroweak force. The theory of this unification is sometimes called QFD (Quantum Flavour Dynamics), or simply GWS (Glashow--Weinberg--Salam). Strictly speaking one can regard the Standard Model as consisting of the theories QFD and QCD. 

The Standard Model is, as far as we know, an exact theory of phenomena accessible at current energies, but some of the mathematics contained within it 
is too complicated and cannot even be solved in a perturbative manner. In QED, the perturbative treatment has been so successful that physicists have claimed the electromagnetic force to be fully disclosed, whereas QCD is not so suitable for perturbative expansions, since it is a highly non-linear theory. For example, the running coupling constant $\alphas$, which is a dimensionless quantity that gives an effective strength of the theory, is larger than unity at low energies or, equivalently, at large distances. In the Feynman perturbative treatment, physical quantities are expanded in powers of $\alphas$, which of course cannot be done for low-energy QCD.  The origin of the perplexity of the strong force comes from the fact that the force carriers, called \emph{gluons}, are massless and carry colour charge themselves, unlike the massive weak bosons and the (electrically) uncharged photons. These two properties mean that the gluons are self-interacting and can be produced without any significant cost of energy.  In turn, this leads to a remarkable property of the strong force called \emph{confinement}. Any quarks found in nature must exist in colour-charge neutral object (mesons and baryons); they are so to speak confined. Simply put, this property implies that there are no free quarks to be observed. To summarize, one can loosely say that QCD is a theory for quarks, but what we also need in order to understand the world is a theory for hadrons, the observable objects in our world. 

Such theories can not be exact (nor unique) since they need to be mathematically simpler than QCD, hence they are called models. The more common ones are `Lattice QCD', bag models and hadronization models. Here we will go into the domain of hadronization models. These are concerned with the process of creating hadrons out of their more fundamental building blocks, the quarks. Models of this kind are of great importance for experimental research, such as high-energy collisions at various accelerator facilities around the world.

The model under consideration in this thesis is the \emph{Lund Hadronization Model}, also called the \emph{Lund String Model} or just the \emph{Lund 
Model}~\cite{string}. The basic assumption of this model is that, because of the confinement property, a colour-field  will mainly stretch out in one space dimension\footnote{This can be compared to the electric field that stretch out in three space dimensions, giving $F=e^2/r^2$ and $V=-e^2/r$}, between a quark and an antiquark in the simplest case, forming a string-like object. The attractive force between the quarks will thereby become constant, denoted $\kappa$, which is equivalent to a linear potential $\kappa r$. We will come back to the more detailed concepts of this model later.
 
Although the Lund Model is a semiclassical theory that only implements Quantum Mechanics in a probabilistic manner (not as a complex amplitude), it has been extremely successful and is widely used throughout the world by experimentalists as well as theorists. However, there are occasions where any simple probabilistic theory breaks down, i.e. when pure quantum mechanical effects manifest themselves. The Bose--Einstein effect is one of these phenomena that were never anticipated by classical physics. Quantum Mechanics requires that a class of particles called \emph{bosons} (integer spin) need to have symmetrized wavefunctions under the exchange of identical particles, whereas the other part of the particle-kingdom, called \emph{fermions} (half-integer spin), has anti-symmetrical wavefunctions under the same type of exchange.  A classical interpretation of this would be that of an attractive force for bosons, and a repulsive force for fermions, acting in both position and momentum space.
  
Nevertheless, the Bose--Einstein effect has been implemented in the Lund Model in a pragmatic way by  L\"onnblad \& Sj\"ostrand~\cite{luboei}. It is taken care of as a local shift in four-momentum between identical mesons according to a two-particle correlation function $c(Q)$, expressing the enhancement in probability to find a pair at a given $Q$, of the form: 
\begin{equation}
c(Q) = 1+\lambda \mathrm{e}^{-(QR)^\eta} \label{first}
\end{equation}
Here $\lambda$ and $R$ are phenomenological parameters, $Q=\sqrt{-(p_1-p_2)^2}$ is the invariant four-momentum difference and $\eta$ is an integer that takes the values 1 or 2, to represent exponential or Gaussian shapes. This form is historically attributed to Hanbury Brown \& Twiss~\cite{hbt} who developed a way to measure star-disk radii using photon interferometry, but later the independent insight of Goldhaber~et~al.~\cite{goldhaber} led to its use in particle physics, as a way to explain the non-isotropic creation of pions in p-$\overline{\p}$ annihilations. The Lund Model results, based on eq.~(\ref{first}), has been verified by experiments for the case of 2-jets, but for higher jet-multiplicity it cannot fully reproduce what is seen. This is given by preliminary analyses of LEP1 data~\cite{kjaer},  which indicate that Bose--Einstein effects are overestimated by the LUBOEI algorithm in 4-jet events, when tuned to 2-jet events. More specifically this method seems to overestimate the Bose--Einstein effect for low energy hadrons. Also theoretical arguments gives that Bose--Einstein effect ought to depend on an `actual' distance $S$ rather than on some typical scale $R$. 

In this thesis I shall try to resolve these problems by associating the constant $R$ with the invariant four-position difference between hadron production vertices, called $S$ (not to be confused with $s$, the squared centre of mass energy), up to a dimensionless constant $k$. For 2-jets there is an approximate correspondence between position and momentum which could explain why we can get by with one argument $(Q)$ in the correlation function. For higher multiplicities of jets the complicated geometry of the string will break that correspondence. What is required is a two-parameter correlation function $c(Q,S)$. We will show that such an approach will reduce the number of hadrons that have non-vanishing Bose--Einstein correlations in higher multiplicities of jets. The effects are not so big, however. Actually the results could also be seen as confirming that the simple ansatz~(\ref{first}) is more meaningful than might have been expected.

The outline of this thesis is as follows: {\bf Section 2} will give a review of the standard model, focusing on the properties of the strong force. In {\bf section 3} we introduce the features of the Lund Hadronization Model, of which the space--time picture and the Bose--Einstein model are our two main areas of interest.   In {\bf section 4} we discuss an extension of the current Lund Bose--Einstein model and in {\bf section 5} the results of the extended model is presented. Finally in {\bf section 6} we summarize and give some ideas that might improve the simple extension for future work in this area. 

\section{The Standard Model}

In this section I will describe the Standard Model in a simplified manner, by showing the known fundamental particles of the Universe and point out some interesting details. I will also try to explain the nature of the interactions by talking a little about groups. The following presentation is neither complete nor easy to follow for those having no prior knowledge, instead I advice~\cite{kane} for a more complete treatment.

\subsection{Matter particles}

The matter particles, six leptons and six quarks, are distinguished by their flavours and are often arranged as doublets in three families or generations:
\[ \begin{array}{cccl}
 \left(\begin{array}{c}\stackrel{\mbox{\large{u}}}{\mbox{\tiny{up}}}\\
  \stackrel{\mbox{\large{d}}}{\mbox{\tiny{down}}}\end{array}\right)&
 \left(\begin{array}{c}\stackrel{\mbox{\large{c}}}{\mbox{\tiny{charm}}}\\
  \stackrel{\mbox{\large{s}}}{\mbox{\tiny{strange}}}\end{array}\right)&
 \left(\begin{array}{c}\stackrel{\mbox{\large{t}}}{\mbox{\tiny{top}}}\\
  \stackrel{\mbox{\large{b}}}{\mbox{\tiny{bottom}}}\end{array}\right) & 
\mbox{quarks} 
\\ \\
 \left(\begin{array}{c}\stackrel{\mbox{\large{$\mathrm{\nu_e}$}}}{\mbox{\tiny{e-
neutrino}}}\\
  \stackrel{\mbox{\large{$\mathrm{e}^-
$}}}{\mbox{\tiny{electron}}}\end{array}\right)&
 \left(\begin{array}{c}\stackrel{\mbox{\large{$\nu_{\mu}$}}}{\mbox{\tiny{$\mu$-
neutrino}}}\\
  \stackrel{\mbox{\large{$\mu^-$}}}{\mbox{\tiny{muon}}}\end{array}\right)&
 
\left(\begin{array}{c}\stackrel{\mbox{\large{$\nu_{\tau}$}}}{\mbox{\tiny{$\tau$-
neutrino}}}\\
  \stackrel{\mbox{\large{$\tau^-$}}}{\mbox{\tiny{tauon}}}\end{array}\right) & 
\mbox{leptons}\\\\
 \mathrm{I} &\mathrm{II} &\mathrm{III} &\mathrm{family/generation}
 \end{array} \]

All the ordinary matter is made up by the first family. The families are similar in many ways but they differ in the particle masses. The third family has the highest masses and the first family has the lowest masses in each row. That is one reason why the families two and three are unstable, except for the neutrinos that cannot decay, only mix. 

The flavours of the matter particles could be said to be different values of one or several quantum numbers that represent mass eigenstates or weak force eigenstates, which are only approximately equal.  The convention however, is to think of them as twelve different particles, because the notion of mass is so important and so very different from other particle properties.  

Besides the flavour, quarks can have three colour charges:  red (r), green (g) and blue (b). The leptons are colourless, or equally well white. Hadrons are also colourless in the sense that they have either all colours $(\mathrm{r}+\mathrm{g}+\mathrm{b}=\mathrm{white}=\mathrm{colourless})$, or they have a colour and an anticolour. The concept of colour in particle physics  is an extended  analogy on Goethe's colour theory. 

Each particle has its corresponding antiparticle (the same symbol with a bar, or the same symbol with reversed sign for charged leptons), which is a particle with the same mass and total spin but with all other properties negated. Since the matter particles are fermions (spin-1/2) they have a helicity quantum number: right-handed particles (index R) have their spin parallel to the direction of motion and left-handed particles (index L) have their spin in the reverse direction. For massive particles, this last property is dependent on the observer, whereas particle/antiparticle is a permanent property to our best knowledge (with some possible exceptions in the neutrino sector).
 
\subsection{Force particles}

The gauge bosons corresponding to the four fundamental forces are:

\[
\begin{array}{llll}
\begin{array}{l}
 \stackrel{\mbox{\normalsize{Particle}}}{\mbox{\tiny{ }}}  \label{tab:force} 
\\\\
\stackrel{\mbox{\normalsize{G}}}{\mbox{\tiny{graviton}}}
\end{array}
&
\begin{array}{l}
\stackrel{\mbox{\normalsize{Interaction}}}{\mbox{\tiny{ }}} \\\\
\stackrel{\mbox{\normalsize{Gravitation}}}{\mbox{\tiny{ }}}
\end{array}
&
\begin{array}{l}
 \stackrel{\mbox{\normalsize{Interacts with}}}{\mbox{\tiny{ }}} \\\\ 
 \stackrel{\mbox{\normalsize{everything}}}{\mbox{\tiny{ }}}
\end{array}
&
\begin{array}{l}
\stackrel{\mbox{\normalsize{Gauge group}}}{\mbox{\tiny{ }}} \\\\
\stackrel{\mbox{\normalsize{ }}}{\mbox{\tiny{ }}}
\end{array}
\\\\
\begin{array}{l}
 \stackrel{\mbox{\normalsize{$\gamma$}}}{\mbox{\tiny{photon}}} \\\\
 \stackrel{\mbox{\normalsize{$\mathrm{W^+,W^-,Z^0}$}}}{\mbox{\tiny{weak 
bosons}}}
\end{array}
&
\begin{array}{l}
\stackrel{\mbox{\normalsize{Electromagnetism}}}{\mbox{\tiny{ }}} \\\\
\stackrel{\mbox{\normalsize{Weak force}}}{\mbox{\tiny{ }}}
\end{array}
&
\begin{array}{l}
\stackrel{\mbox{\normalsize{electric 
charge}}}{\mbox{\tiny{(quarks,$\mathrm{e,\mu,\tau,W^{\pm}}$)}}} \\\\
\stackrel{\mbox{\normalsize{weak isospin 
charge}}}{\mbox{\tiny{(quarks,leptons,$\mathrm{W^{\pm},Z^0}$)}}}
\end{array}
&
\left.
\begin{array}{l}
\stackrel{\mbox{\normalsize{$\mathrm{U_{em} (1)}$}}}{\mbox{\tiny{ }}} \\\\
 \stackrel{\mbox{\normalsize{ }}}{\mbox{\tiny{ }}}
\end{array}\right\}\mathrm{U_Y (1) \times SU_L (2)} \\\\
\begin{array}{l}
 \stackrel{\mbox{\normalsize{$\mathrm{g}_{c\bar{c}'}$}}}{\mbox{\tiny{8 gluons}}} 
\\\\
\end{array}
&
\begin{array}{l}
\stackrel{\mbox{\normalsize{Strong Force}}}{\mbox{\tiny{ }}} \\\\
\end{array}
&
\begin{array}{l}
 \stackrel{\mbox{\normalsize{colour charge}}}{\mbox{\tiny{(quarks, gluons)}}} 
\\\\
\end{array}
&
\begin{array}{l}
\stackrel{\mbox{\normalsize{$\mathrm{SU_c (3)}$}}}{\mbox{\tiny{ }}} \\\\
 \end{array} 
\end{array}
\]

The graviton is not really predicted by the Standard Model but is included here for the sake of completeness. The photon is the carrier of the electromagnetic force and is its own antiparticle. The weak force has three gauge bosons $\mathrm{\W^+,\W^-}$ and $\mathrm{Z^0}$, where the W's are each other's 
antiparticles and the $\mathrm{Z^0}$ is its own antiparticle. Finally there are eight gluons carrying different combinations of colour charges, denoted by indices $c\bar{c}'$ in the table above.

All the force particles are bosons and thus have integer spin; the graviton is a spin-2 particle, whereas the others are spin-1 particles. 

The gauge bosons of the weak force are special since they have non-zero masses. In fact they are surprisingly heavy; the W's are around 86, and the $\mathrm{Z^0}$ is about 97, times heavier than the Proton (938 MeV). Because of the Heisenberg uncertainty principle $\Delta t \Delta E \sim \hbar$ the range of the weak force is limited, thereby making it weak.

What the charge of the weak force is might not be intuitively clear. In the table of the matter particles  the parentheses are supposed to mark out weak iso-spinors, of which the single particles are merely projections. In analogy to the well-known theory of spin, particles in the upper state of an iso-spinor has a third component of the weak charge being $+1/2$, and particles in the lower state has $-1/2$. This is however only true for left-handed particles; it is a strange fact that all right-handed particles have no iso-spin charge, i.e. they are singlets.

Like the left-handed matter particles are doublets in an intrinsic iso-spin space, quarks are triplets in a colour space. In this language different 
interactions corresponds to orthogonal rotations in the intrinsic spaces; such rotations belong to mathematical groups  called SU($n$) if $n>1$, and U(1) if the dimension equals 1. These are also called gauge groups, and in the Standard Model one usually give them an index that indicates which interaction the groups correspond to.

As seen in the above table, the weak force, defined by the interaction particles $\mathrm{W^+,W^-,Z^0}$), does not have an independent gauge group associated to it. The weak force is embedded in a larger gauge group called  $\mathrm{U_Y (1) \times SU_L (2)}$, which is the electroweak gauge group that also contains the electromagnetic group.
 
\subsection{Quantum Chromo-Dynamics} 

In this thesis our main interest lies in the nature of hadronization, the process that creates hadrons out of the more fundamental quarks. This process is controlled by the strong force; it is included at the fundamental level in the QCD framework, but in reality it is less well understood. Nevertheless, I will state the basic equation of QCD and discuss some interesting properties of the strong force.

\subsubsection{The QCD Lagrangian}

The Lagrangian is in principle made out of the sum of  Dirac\footnote{The Dirac Lagrangian give rise to the Dirac equation, which is the correct relativistic generalization of the Schr\"odinger wave equation for fermions}  Lagrangians of all known quarks and a Lagrangian term for the gluons. 
\begin{equation}
\mathcal{L}_\mathrm{QCD}= 
\sum_{q=\mathrm{u,d,c,s,t,b}}\bar{q}i\gamma^{\mu}\mathcal{D}_{\mu}q -
\frac{1}{2}\mathrm{tr}(G^{\mu\nu}G_{\mu\nu})
\end{equation}
The last term is very similar to the Lagrangian of the electromagnetic field. The symbols for the quarks should be understood as wavefunctions, or (spin-)vector fields, and the eight gluons are combined into one matrix field $G_\mu$, all parameterized by space--time co-ordinates $x^\mu=(t,x,y,z)$.

The covariant derivative, that substitutes the ordinary derivative, includes a gluon field term that causes the quarks and gluons to interact.
\begin{equation}
\mathcal{D}_{\mu}= \partial_{\mu}-ig_3 G_{\mu} 
\end{equation}
The field strength tensor of the gluon field is a SU(3) generalization of the electromagnetic field strength tensor; thus it is a 3 by 3 matrix in colour space.
\begin{equation}
G_{\mu\nu}=\partial_{\mu}G_{\nu}-\partial_{\nu}G_{\mu}-i2g_3[G_{\nu},G_{\mu}]
\end{equation}
The commutator term in $G_{\mu\nu}$ contains the physics that makes the strong force extraordinary compared to electromagnetism. It arises because SU(3) is a non-Abelian group (elements do not commute in general). It is an easy matter to show that, when the gluon field term of the Lagrangian is expanded, one obtains cubic and quartic terms in $G_\mu$, that are not there for an Abelian group, such as electromagnetism.

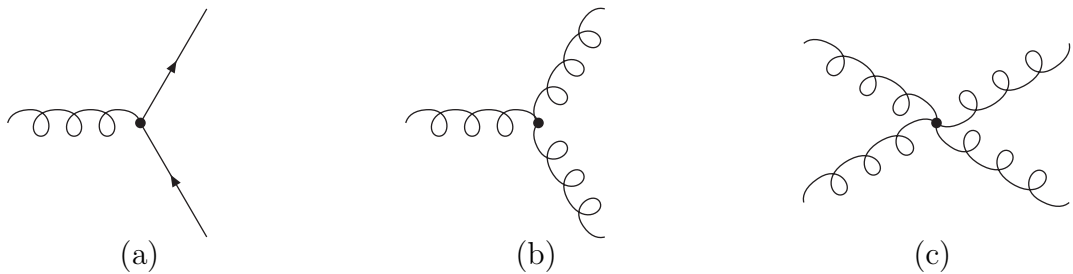
\begin{figure} \begin{center} \begin{picture}(500,120)(0,30)
\Vertex(100,100){2}
\Text(100,50)[]{(a)}
\Gluon(50,100)(100,100){5}{3}
\ArrowLine(125,57)(100,100)
\ArrowLine(100,100)(125,143)
\Vertex(250,100){2}
\Text(250,50)[]{(b)}
\Gluon(200,100)(250,100){5}{3}
\Gluon(275,57)(250,100){5}{3}
\Gluon(250,100)(275,143){5}{3}
\Vertex(400,100){2}
\Text(400,50)[]{(c)}
\Gluon(350,130)(400,100){5}{3}
\Gluon(350,70)(400,100){5}{3}
\Gluon(400,100)(450,130){-5}{3}
\Gluon(400,100)(450,70){-5}{3}
\end{picture} \caption{Different types of interactions in QCD: (a) 2 quarks and 
a gluon, (b) 3 gluons, (c) 4 gluons.} \label{fig:vertex}
\end{center}
\end{figure}

The field equations of the QCD Lagrangian are unsolvable. Especially complicated is the equation for the gluon field, which is cubic in $G_\mu$ (one order lower than the Lagrangian).  In practice, at high energies, one can use the Feynman rules to draw diagrams with vertices and propagators, and do perturbative calculations. Each term in the Lagrangian with more than two fields is called an interaction term, which corresponds to a diagram where the particles meet in one vertex with a coupling strength as indicated by the pre-factors of the term in question.  From the QCD Lagrangian  it is clear that there exist three classes of interaction vertices.  The first one is when a gluon couples to two quarks of the same kind. The `same kind' can here also mean a quark and an antiquark of the same kind, depending on which direction you choose the time-axis in\footnote{An antiparticle is equivalent to a particle travelling backwards in time.}. The second vertex is three gluons interacting, and the third one is four gluons interacting, as shown in Fig.~\ref{fig:vertex}. 

At low energies the Feynman rules are not applicable to QCD, instead one has to approach phenomenological rules to obtain a decent description of the physics.  
 
\subsubsection{Confinement}

Just as charge neutral composite particles are energetically favoured by electromagnetism so are colourless particles favoured in the theory of QCD. Although it has not been proven from first principles, we have strong reasons to believe that the potential barrier to extract a free quark is infinitely high. The confinement property thus ensures that no quarks (colour triplets) or  gluons (octets) can exist as free particles; they can only exist within mesons and baryons/antibaryons (all singlets). 

A meson consists of an antiquark with anticolour and a quark with colour. A baryon consists of three quarks, all with different colours.  An antibaryon is then a combination of three differently (anti-)coloured antiquarks. The wave functions of the baryons/mesons has to be anti-symmetrized/symmetrized with respect to exchange of colours, so nothing precise can be said about what colour a single quark has.  

From an experimental point of view the confinement property is a hassle, because it hides the underlying physics from our eyes. But with modern high-energy accelerators some information on confined particles reveals itself in the form of jets. Jets are showers of collimated hadrons that arise after particle collisions. The formation of jets is a proof that hadrons consist of quarks, and especially for $n$-jets with $n$ larger than two it is the best proof of the existence of the gluon.  

For example, a collision that takes place between two protons primarily involves a few quarks and maybe some gluons. These particles will violently accelerate and change their direction of propagation, which is unmeasurable to us since this happens inside a confined `bubble'. But the bubble will break into pieces, and form hadrons that inherit the line of propagation from the accelerated quarks and gluons.  Thus we see jets in our detectors. 

\section{The Lund Hadronization Model}

Experimental as well as theoretical work has shown that the attractive force between two oppositely coloured particles is essentially independent of the distance at large separations. This fact is believed to have its origins in the non-linear gluon--gluon interactions in the surrounding colour field. The interactions cause the field lines to compactify into a flux tube (or string) of constant radius in order to minimize the action. One can associate a linear energy density $\kappa$ with the flux tube. More commonly we refer to $\kappa$ as the string tension, which is measured to have the pleasant value of $\simeq 1 \mathrm{\ GeV/fm}$. 

It is possible to use this fact to extract some physics, as it is done in the Lund Model. One could say that the basic ingredients to produce a feasible model that describes hadronization is: 
\begin{itemize}
\item A classical linear potential $V(r)=\kappa r$. 
\item A `matter' string, on which gluons are distributed, and with a quark and 
an antiquark attached at the two end points.  
\item A semiclassical fragmentation model, approximating the correct Quantum 
Mechanical picture, such that it can be implemented with Monte Carlo methods.
\end{itemize}

From these principles, in addition to several unmentioned,  it is possible to consider some general classes of events that occur after particle collisions at high energies. In this study we restrict our attention to $\mathrm{e^+e^-}$ annihilation events. Then the simplest process is the creation of  a quark--antiquark pair, $\mathrm{e^+e^-} \rightarrow \mathrm{\gamma/Z^0} \rightarrow \mathrm{q\bar{q}}$. This is the two-parton case, which could for the moment also be called the two-jet case, but we wish to reserve this name for experimentally measured events; this difference in terminology will become clear later.  The name \emph{parton} historically applies to both quarks/antiquarks and gluons. Later it is worth discussing the  3-parton case, which will contribute with some interesting physics, that is used to extend our picture into a general $n$-parton case. 

\subsection{The 2-parton case}\label{sec:2part}

\begin{figure}\begin{center}\begin{picture}(400,200)(-100,0)
\Vertex(-70,30){2}
\Vertex(-70,60){2}
\Photon(-70,30)(-70,60){5}{3}
\ArrowLine(-100,0)(-70,30)
\ArrowLine(-40,0)(-70,30)
\Text(-105,5)[]{$\mathrm{e^+}$}
\Text(-30,5)[]{$\mathrm{e^-}$}
\Text(-105,85)[]{$\mathrm{\bar{q}}$}
\Text(-35,85)[]{$\mathrm{q}$}
\Text(-70,80)[]{\scriptsize String}
\Text(-55,60)[]{$E_\mathrm{cm}$}
\ArrowLine(-70,60)(-100,90)
\ArrowLine(-70,60)(-40,90)
\DashLine(-100,90)(-40,150){3}
\DashLine(-40,90)(-100,150){3}
\DashLine(-40,150)(-100,210){3}
\DashLine(-100,150)(-40,210){3}

\SetOffset(100,-80)

\rotatebox{45}{
\SetScale{0.80}
\SetWidth{0}
\GBox(0,0)(200,4){0.8}
\GBox(0,0)(104,12){0.8}
\GBox(0,0)(19,115){0.8}
\GBox(0,0)(4,200){0.8}
\GBox(0,0)(29,71){0.8}
\GBox(0,0)(59,35){0.8}
\SetWidth{0.5}
\LongArrow(0,0)(0,200)
\LongArrow(0,0)(200,0)
\Line(104,4)(200,4)\Line(200,0)(200,4)
\DashLine(104,0)(104,4){2}
\GBox(200,4)(296,8){0.8}
\Line(59,12)(104,12)\Line(104,4)(104,12)
\DashLine(104,4)(59,4){2}
\DashLine(59,4)(59,12){2}
\GBox(104,12)(149,20){0.8}
\GBox(149,20)(194,28){0.8}
\Line(29,35)(59,35)\Line(59,12)(59,35)
\DashLine(59,12)(29,12){2}
\DashLine(29,12)(29,35){2}
\GBox(59,35)(89,58){0.8}
\GBox(89,58)(119,82){0.8}
\GBox(119,82)(149,105){0.8}
\LongArrow(119,82)(149,105)
\Line(19,71)(29,71)\Line(29,35)(29,71)
\DashLine(29,35)(19,35){2}
\DashLine(19,35)(19,71){2}
\GBox(29,71)(39,107){0.8}
\GBox(39,107)(49,143){0.8}
\GBox(49,143)(59,179){0.8}
\Line(4,115)(19,115)\Line(19,71)(19,115)
\DashLine(19,71)(4,71){2}
\DashLine(4,71)(4,115){2}
\GBox(19,115)(34,159){0.8}
\GBox(34,159)(49,203){0.8}
\Line(4,115)(4,200)\Line(0,200)(4,200)
\DashLine(4,115)(0,115){2}
\GBox(4,200)(8,285){0.8}
\SetScale{1}}
\SetOffset(43,-80)
\Text(145,190)[]{$k_0$}
\Text(-85,190)[]{$k'_0$}
\Text(24,143)[]{\scriptsize $\mathrm{\bar{q}}_i \mathrm{q}_i$}
\Text(85,125)[]{\scriptsize $\mathrm{\bar{q}}_{i-1} \mathrm{q}_{i-1}$}
\Line(24,140)(24,132)
\Line(85,129)(55,134)

\SetOffset(100,-80)

\Text(-15,240)[]{$p_i$}
\Text(-9,193)[]{$m_i^2$}
\Text(-28,111)[]{\scriptsize String}

\LongArrow(80,90)(80,130)
\LongArrow(80,90)(120,90)
\Text(87,130)[]{$\hat{t}$}
\Text(133,90)[]{$\hat{x}$}
\end{picture}  \caption{Left: A typical event that give rise to a $\mathrm{q\bar{q}}$-pair. Right: A zoomed in picture of the string as it fragments into hadrons. Note: space--time and energy-momentum co-ordinates are used interchangeably} \label{fig:2jet}
\end{center}\end{figure}
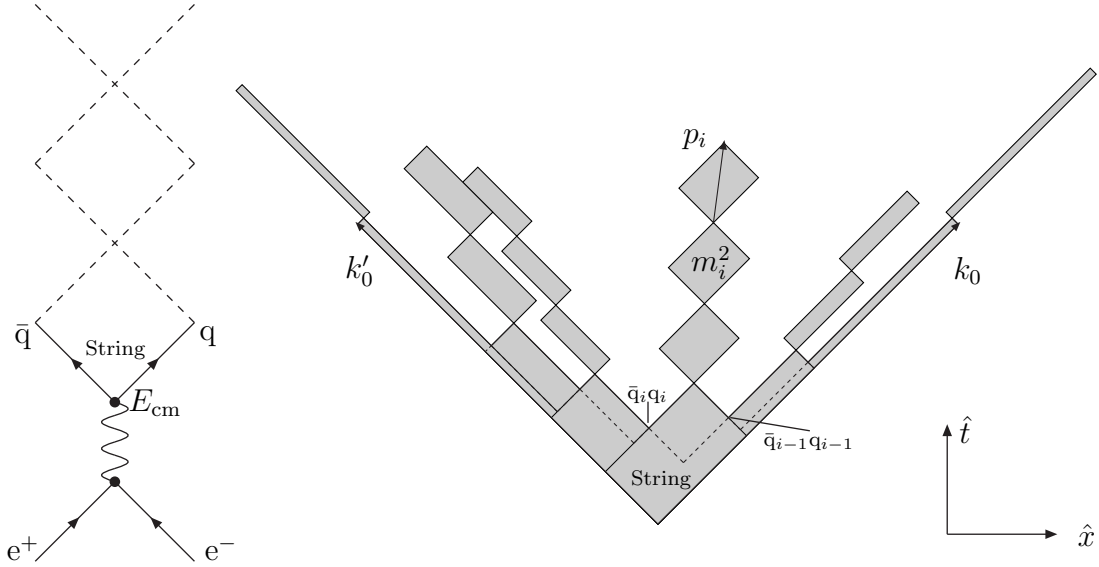

Assume we study an $\mathrm{e^-e^+}$ annihilating collision in its rest frame, with centre of mass energy $E_\mathrm{cm}$, that creates a quark/antiquark-pair ($\mathrm{q\bar{q}}$). Assume further that the quarks we are discussing on the following pages are massless. The quarks will go out back-to-back (thus defining an $\hat{x}$-axis) with the speed of light $(c=1)$, spanning a gluon field string in-between. If we for a moment imagine that the string cannot break, the $\mathrm{q\bar{q}}$-pair will run out of energy when it reaches a separation of $E_\mathrm{cm}/\kappa$, and will be forced to turn back into a oscillating mode, as indicated by the dashed lines in the left part of Fig.~\ref{fig:2jet}. Since we have a linear potential and massless quarks, there is a direct way to convert between the energy--momentum (in GeV) of a quark and the straight sections of its space--time world-line (in fm), with the help of $\kappa=1\mathrm{\ GeV/fm}$. Thus the 4-momentum and the 4-position can neatly be visualized in the same diagram.

In reality, for the collisions studied in the Lund Model,  $E_\mathrm{cm} \gg m_\mathrm{hadron}$ for any metastable hadron. Thus the string will break up into smaller fractions even before the original quarks reach maximum separation. The breaking up is caused by the production of virtual $\mathrm{q\bar{q}}$-pairs that acquire 4-momentum from the string, allowing them to become real and screen off the colour of the two end-quarks. In principle a new $\mathrm{q\bar{q}}$-pairs could be produced anywhere in the string, but the breaks have to be correlated such that the hadron produced between two breaks has the correct physical mass and such that the total energy and momentum of the system is conserved in the fragmentation process. In the Lund Model this is taken care by an iterative Monte Carlo algorithm, that could be sketched in the following way.

It is assumed that $\mathrm{\bar{q}}_i \mathrm{q}_i$-pairs are created, where $i$, counted from right to left, runs from 1 to $n-1$, $n$ being number of hadrons. Hadron $i$ is then made up of the quarks $\mathrm{\bar{q}}_{i} \mathrm{q}_{i-1}$, where $\mathrm{\bar{q}}_{n}$ and $ \mathrm{q}_0$ are the two original quarks.

If we let $k_0$ denote the 4-momenta of $\mathrm{q}_0$ and $k'_0$ that of $\mathrm{\bar{q}}_{n}$, then the first hadron, with 4-momentum $p_1$ and mass $m_1$, uses a fraction $z_1$ of $k_0$ when it is created. $z_i$ is a random variable with probability density function 
\begin{equation}
f(z) \propto \frac{1}{z}(1-z)^a \mathrm{e}^{-\frac{bm_i^2}{z}} \label{zed}
\end{equation}
where $a$ and $b$ are phenomenological constants; $a \approx 1$ and $b \approx 0.5 \ \mathrm{GeV}^{-2}$ . Also the hadron uses a fraction $z'_1$ out of  $k'_0$, where $z'_1$ is fixed by the restriction $p_1^2=(z_1 k_0+z'_1 k'_0)^2=m_1^2$ to:
\begin{equation}
z'_1=\frac{m_1^2}{2z_1 k_0\cdot k'_0}
\end{equation}
The second hadron ($p_2$ and $m_2$) will have the remaining $k_1=(1-z_1)k_0$ and $k'_1=(1-z'_1)k'_0$ at its disposal. These 4-momenta substitute  $k_0$ and $k'_0$ above; other than that, the reasoning is the same as for the first hadron. Thus one can define the remaining 4-momentum $k_i$, after $i$ hadrons have been produced,  to be:
\begin{equation}
k_i=\prod_{j=1}^i (1-z_j)k_0
\end{equation}
Here, the same equation applies to $k'_i$ if one primes $k_0$ and $z_j$.

The iterative procedure outlined above is easily continued any number of steps, so that every hadron $(p_i,m_i)$ has its associated fractions $z_i$ and $z'_i$. It can be continued as long as the remaining 4-momenta allows more hadrons to be produced; this gives us the number of hadrons $n$, which will vary from event to event. The constraints coming from overall energy and momentum conservation implies that not all $z_i$ can be chosen independently according to eq.~(\ref{zed}): the last two hadrons are necessarily fixed by the constraints. To minimize any non-physical effects of the choice of `last two', fragmentation processes are allowed to begin in the two ends of the string and meet up at some point that is randomly selected.  

In the space--time picture (Fig.~\ref{fig:2jet}) one can calculate the vertex where a  $\mathrm{q}_i \mathrm{\bar{q}}_i$-pair is created. If the vertices are counted from the upper right corner, then $w_i=k_i/\kappa+(k'_0-k'_i)/\kappa$ is the vector of the $i$th vertex. [Throughout this thesis we will consistently use the notation $w_i$ for break-up vertices ($\mathrm{q}_i \mathrm{\bar{q}}_i$-production points), and later $v_i$'s will denote hadron production vertices.] From Fig.~\ref{fig:2jet} it is clear that hadrons are represented by rectangles with areas equal to their mass-square, and with time-like diagonals equal to their 4-momenta $p_i$. When created, each hadron is connected to two neighbouring hadrons through the $\mathrm{q\bar{q}}$-vertices that define it. Of course, the hadronization picture used here is somewhat too ambitious, because the Heisenberg uncertainty principle does not allow us to speak simultaneously about the momentum and the position of the quarks inside a hadron. Thus the space--time picture should not be taken too literally; it is primarily used in the Lund Model as a way to deduce the 4-momenta of the hadrons in a particular event. Later we will try to expand the interpretation somewhat, such that one might speak of an approximate creation position of a hadron, which would be represented by some point inside the dashed rectangles in the right part of Fig.~\ref{fig:2jet}.
 
To make the model more realistic, a transverse momentum is added to each quark, with a Gaussian distribution $p_\perp \sim 300$ MeV, in a manner such that it vanishes for each $q\bar{q}$-pair originating from the same vertex. As a consequence we have to replace $m^2$ by the transverse mass $m_\perp^2=m^2+p^2_\perp$, and all the 4-vectors used have to be projected onto the plane spanned by the two original 4-momenta $k_0$ and $k'_0$, everywhere in the above equations and arguments.

\subsection{The 3-parton case} \label{sec:3jet}

If we have a three-particle system, e.g. created by $\mathrm{e^+e^- \rightarrow \gamma/Z^0 \rightarrow q\bar{q}g}$, it will in general spread out in two space dimensions (the $\hat{x}\hat{y}$-plane). In the Lund Model this is represented by a string with a kink on it. The kink/gluon carries 4-momentum, colours and spin, of which the first is central in the following discussion whereas the last property is neglected. The actual colours of the gluon are not very important, but for the sake of our discussion we may assign colours to the partons according to $\mathrm{q=r}$, $\mathrm{g=\bar{r}b}$, $\mathrm{\bar{q}=\bar{b}}$; this implies that the string flows from q to g and onto $\mathrm{\bar{q}}$.

The kinematics of the string is more easily visualized if one boosts a three-parton system such that the momentum directions of the quark and antiquark is parallel respectively antiparallel to the $\hat{x}$-axis, and the gluon momentum is parallel  to the $\hat{y}$-axis. 

\begin{figure} \begin{picture}(600,360)(80,-290)
\SetOffset(160,90)
\GBox(25,-80)(270,-10){0.8}
\LongArrow(100,-50)(50,-50) \Text(58,-60)[]{$\vec{p}_\mathrm{\bar{q}}$} 
\LongArrow(100,-50)(175,-50) \Text(170,-60)[]{$\vec{p}_\mathrm{q}$} 
\LongArrow(100,-50)(100,-25) \Text(110,-30)[]{$\frac{\vec{p}_\mathrm{g}}{2}$}
\Vertex(100,-50){2}
\Text(100,-65)[]{$t=0$}
\LongArrow(220,-50)(220,-25)
\Text(228,-25)[]{$\hat{y}$}
\LongArrow(220,-50)(245,-50)
\Text(253,-50)[]{$\hat{x}$}

\SetOffset(65,10)

\DashLine(100,-50)(50,-50){2} \LongArrow(55,-50)(50,-50) \Text(70,-60)[]{$\bar{\q}$} \Text(69,-51)[]{$\leftarrow$}
\DashLine(100,-50)(175,-50){2} \LongArrow(170,-50)(175,-50) \Text(130,-60)[]{q} 
\Text(131,-51)[]{$\rightarrow$}
\DashLine(100,-50)(100,-25){2} \LongArrow(100,-30)(100,-25) \Text(105,-15)[]{g}
\Line(75,-50)(100,-25) \Line(100,-25)(125,-50) \Vertex(75,-50){2} \Vertex(100,-25){2} \Vertex(125,-50){2}
\Text(100,-65)[]{$t=1$}
\DashLine(230,-50)(180,-50){2} \LongArrow(185,-50)(180,-50) \Text(190,-60)[]{$\bar{\q}$}
\DashLine(230,-50)(305,-50){2} \LongArrow(300,-50)(305,-50)\Text(290,-60)[]{q} 
\Text(286,-51)[]{$\rightarrow$}
\DashLine(230,-50)(230,-25){2} \LongArrow(230,-30)(230,-25)\Vertex(180,-50){2} 
\Vertex(280,-50){2}
\Line(180,-50)(205,-25) \Line(205,-25)(255,-25) \Line(255,-25)(280,-50)
\Text(230,-65)[]{$t=2$}
\DashLine(360,-50)(310,-50){2} \LongArrow(315,-50)(310,-50) \Text(320,-35)[]{$\bar{\q}$}
\DashLine(360,-50)(435,-50){2} \LongArrow(430,-50)(435,-50) \Text(430,-60)[]{q}
\DashLine(360,-50)(360,-25){2} \LongArrow(360,-30)(360,-25) 
\Vertex(310,-25){2} \Text(312,-20)[]{$\uparrow$}\Line(312,-25)(410,-25) 
\Line(410,-25)(435,-50) \Vertex(435,-50){2}
\Text(360,-65)[]{$t=3$}

\SetOffset(-326,-90)

\DashLine(490,-50)(440,-50){2} \LongArrow(445,-50)(440,-50) \Text(440,-10)[]{$\bar{\q}$}
\DashLine(490,-50)(565,-50){2} \LongArrow(560,-50)(565,-50) \Text(560,-35)[]{q}
\DashLine(490,-50)(490,-25){2} \LongArrow(490,-30)(490,-25) 
\Vertex(440,0){2} \Line(440,0)(465,-25) \Line(465,-25)(565,-25) \Vertex(565,-25){2}\Text(567,-20)[]{$\uparrow$}
\Text(490,-65)[]{$t=4$}

\SetOffset(195,40)

\DashLine(100,-180)(50,-180){2} \LongArrow(55,-180)(50,-180) \Text(75,-140)[]{$\bar{\q}$} \Text(81,-131)[]{$\rightarrow$}
\DashLine(100,-180)(175,-180){2} \LongArrow(170,-180)(175,-180) \Text(175,-140)[]{q}
\DashLine(100,-180)(100,-155){2} \LongArrow(100,-160)(100,-155) 
\Line(75,-130)(100,-155) \Line(100,-155)(150,-155) \Line(150,-155)(175,-130)
\Vertex(75,-130){2} \Vertex(175,-130){2}
\Text(100,-195)[]{$t=5$}
\DashLine(230,-180)(180,-180){2} \LongArrow(185,-180)(180,-180) \Text(236,-131)[]{$\rightarrow$}
\DashLine(230,-180)(305,-180){2} \LongArrow(300,-180)(305,-180) \Text(275,-131)[]{$\leftarrow$}
\DashLine(230,-180)(230,-155){2} \LongArrow(230,-160)(230,-155) 
\Line(230,-130)(255,-155) \Line(255,-155)(280,-130) \Text(225,-140)[]{$\bar{\q}$} \Text(260,-165)[]{g}
\Vertex(230,-130){2} \Vertex(255,-155){2} \Vertex(280,-130){2} \Text(285,-140)[]{q}
\Text(230,-195)[]{$t=6$}

\SetOffset(-130,-80)

\DashLine(360,-180)(310,-180){2} \LongArrow(315,-180)(310,-180)\Text(395,-140)[]{$\bar{\q}$}
\DashLine(360,-180)(435,-180){2} \LongArrow(430,-180)(435,-180) \Text(375,-140)[]{q}
\DashLine(360,-180)(360,-155){2}  \LongArrow(360,-160)(360,-155) 
\Text(393,-120)[]{g}
\Vertex(385,-130){2} \Text(387,-126)[]{$\uparrow$} \Text(391,-132)[]{$\rightarrow$} \Text(380,-132)[]{$\leftarrow$}
\Text(360,-195)[]{$t=7$}
\DashLine(490,-180)(440,-180){2} \LongArrow(445,-180)(440,-180) \Text(545,-140)[]{$\bar{\q}$} \Text(546,-131)[]{$\rightarrow$}
\DashLine(490,-180)(565,-180){2} \LongArrow(560,-180)(565,-180) \Text(485,-140)[]{q} \Text(485,-131)[]{$\leftarrow$}
\DashLine(490,-180)(490,-155){2} \LongArrow(490,-160)(490,-155) \Text(520,-95)[]{g}
\Line(490,-130)(515,-105) \Line(515,-105)(540,-130) \Vertex(490,-130){2} 
\Vertex(515,-105){2} \Vertex(540,-130){2}
\Text(490,-195)[]{$t=8$}
\end{picture} \caption{Snapshots of the string at different times.} 
\label{fig:space}
\end{figure}
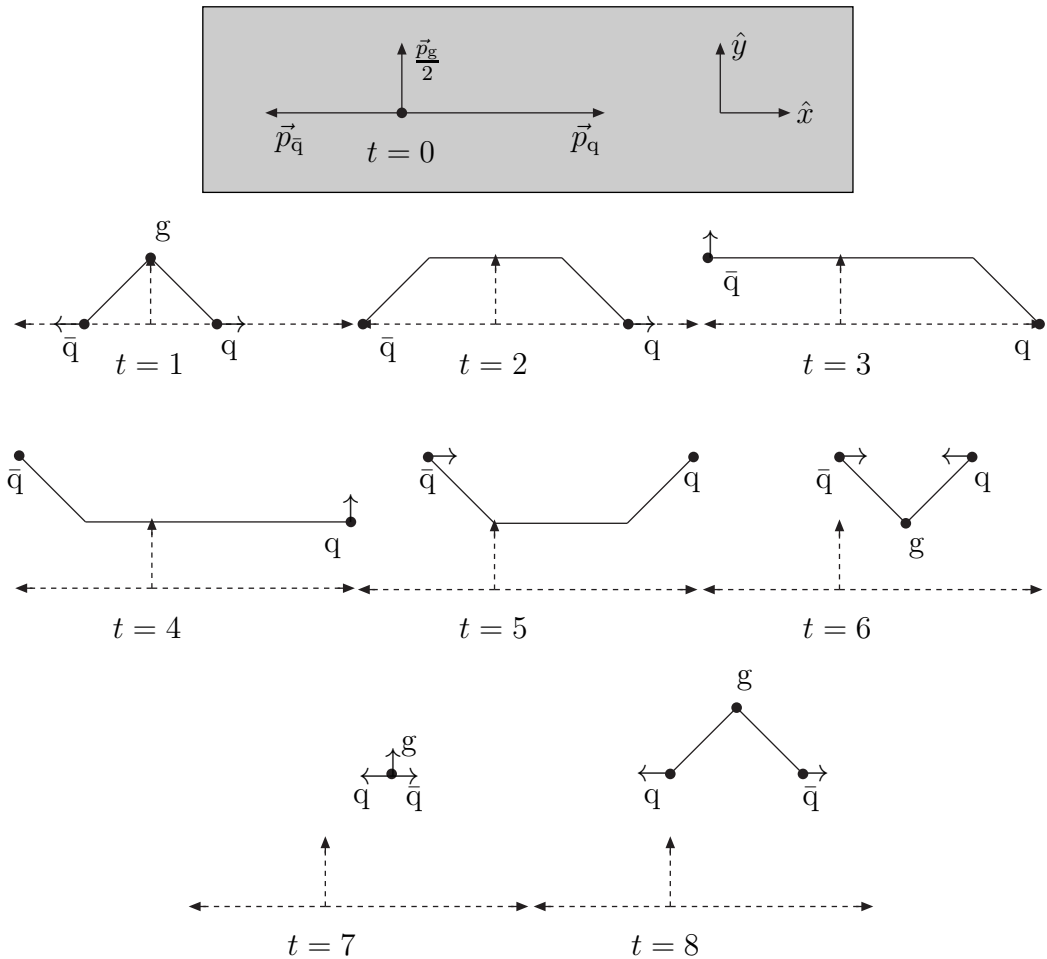 
As seen in Fig.~\ref{fig:space}, the particles will travel in their own directions unaffected by what is going on at other point of the string. Meanwhile, they lose energy and momentum to the string being pulled out by them. Eventually the particles will run out of energy. Here, we assume that this happens to the gluon first, which is normally the case, but this does not do the argumentation less general. The gluon stops after travelling a distance corresponding to half its energy: since it has two string-ends attached to it, it loses energy twice as fast as a quark. As a consequence a third string region is formed that carries no momentum; it has only energy in this choice of reference frame. There are also two kinks formed when the gluon stops; they are however different from a gluon kink, since these two carry only information about the change of motion and do not have a local 4-momentum concentration that is usually associated with particles.

Later the quarks will also run out of energy, forcing them to stop. Then the virtual gluons in the string will finally catch up and make their presence felt. Since the string pieces closest to the quarks have a transverse momentum inherited from the gluon, the quarks start to move in the direction of the original gluon-momentum. The quarks will eat up a piece of the string and gain energy--momentum until they reach a point where the string is left behind again. The process repeats itself once more: a deceleration followed by a stop and change of path, followed by acceleration. Eventually all the three particles will meet in one single point translated away from the origin.  This corresponds to a half cycle of the gluon--quark--antiquark system at $t=7$ in Fig.~\ref{fig:space}. The total momentum of the system exactly corresponds to the translated distance, which can easily be seen by adding up the original momenta that are indicated with dashed arrows (only half of the gluon momentum is showed). 

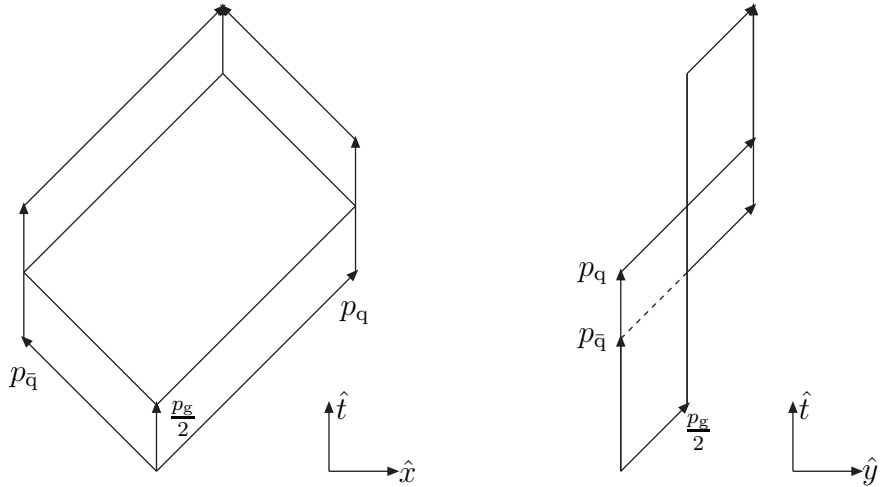
\begin{figure}   \begin{picture}(250,200)(-80,0)
\LongArrow(0,0)(75,75)
\LongArrow(0,0)(-50,50)
\LongArrow(0,0)(0,25)
\LongArrow(75,75)(75,125)
\LongArrow(-50,50)(-50,100)
\Line(0,25)(75,100)
\Line(0,25)(-50,75)
\Line(75,100)(25,150)
\Line(-50,75)(25,150)
\LongArrow(25,150)(25,175)
\LongArrow(75,125)(25,175)
\LongArrow(-50,100)(25,175)
\Text(75,60)[]{$p_\mathrm{q}$}
\Text(10,20)[]{$\frac{p_\mathrm{g}}{2}$}
\Text(-50,35)[]{$p_\mathrm{\bar{q}}$}
\SetOffset(15,0)
\LongArrow(50,0)(50,25)
\Text(55,25)[]{$\hat{t}$}
\LongArrow(50,0)(75,0)
\Text(80,0)[]{$\hat{x}$}
\SetOffset(25,0)
\LongArrow(150,0)(150,75)
\LongArrow(150,0)(150,50)
\LongArrow(150,0)(175,25)
\LongArrow(150,75)(200,125)
\DashLine(150,50)(175,75){2}
\LongArrow(175,75)(200,100)
\Line(175,25)(175,100)
\Line(175,25)(175,75)
\Line(175,100)(175,150)
\Line(175,75)(175,150)
\LongArrow(175,150)(200,175)
\LongArrow(200,125)(200,175)
\LongArrow(200,100)(200,175)
\Text(140,75)[]{$p_\mathrm{q}$}
\Text(180,15)[]{$\frac{p_\mathrm{g}}{2}$}
\Text(140,50)[]{$p_\mathrm{\bar{q}}$}
\SetOffset(40,0)
\LongArrow(200,0)(200,25)
\Text(205,25)[]{$\hat{t}$}
\LongArrow(200,0)(225,0)
\Text(230,0)[]{$\hat{y}$}
\end{picture}
\caption{Projections of space--time sheet. The right picture is rotated 90 degrees in the $\hat{x}\hat{y}$-plane, with respect to the left picture. The arrows represent the parton world lines, the plain lines represent non-gluon kink world lines and the dashed line indicate a hidden part of the sheet.} 
\label{fig:diamond}
\end{figure}

In this choice of reference frame it looks like a coincidence that the partons and the kinks travel in directions always parallel to the original momentum of the three particles, but this is of course true in all frames (for massless partons). In fact the space--time world lines of the partons and the kinks can be built up by shorter lines that correspond to these three original 4-momentum vectors. The half cycle space--time sheet of the string will then look like Fig.~\ref{fig:diamond}. There are five different regions depending on which pair of four-vectors that spans them.  They all have a diamond shape and can separately be boosted into their own rest frames, such that one obtains something that looks like the 2-parton case.
  
As in the 2-parton case we consider only systems with high energy, which are the ones where the string model is valid. The 3-parton system will thus fragment into smaller pieces, corresponding to hadrons, at an early stage and will never evolve as far as Fig.~\ref{fig:diamond} shows. The fragmentation process is essentially the same as in the 2-parton case for the two lower regions. In the middle region the available 4-momentum is not defined by the two vectors that span it, but by the remaining fractions of these vectors that is not used by hadrons in lower regions, thus reducing the number of hadrons that can produced with the fragmentation algorithm in this region. All fragmentation in later regions can be neglected in this treatment.

\begin{figure}  \begin{center} \begin{picture}(200,200)(-100,-200)
\Text(123,-185)[]{$p_\mathrm{q}$}
\Text(-20,-185)[]{$\frac{p_\mathrm{g}}{2}$}
\Text(35,-185)[]{$\frac{p_\mathrm{g}}{2}$}
\Text(-105,-185)[]{$p_\mathrm{\bar{q}}$}
\Text(47,-115)[]{$p_\mathrm{q}$}
\Text(-30,-115)[]{$p_\mathrm{\bar{q}}$}
\rotatebox{-135}{
\Line(190,0)(190,40)
\DashLine(200,40)(190,40){2}
\Line(190,40)(175,40)
\Line(175,40)(175,60)
\DashLine(190,40)(190,60){2}
\DashLine(190,60)(175,60){2}
\Line(175,60)(150,60)
\Line(150,60)(150,75)
\DashLine(175,60)(175,75){2}
\DashLine(175,75)(150,75){2}
\Line(150,75)(120,75)
\Line(120,75)(120,85)
\DashLine(150,75)(150,85){2}
\DashLine(150,85)(120,85){2}
\Line(120,85)(100,85)
\Line(100,85)(95,85)
\Line(95,85)(95,92)
\DashLine(120,85)(120,92){2}
\DashLine(120,92)(95,92){2}
\Line(95,92)(85,92)
\Line(85,92)(85,100)
\Line(85,100)(85,115)
\DashLine(95,92)(95,115){2}
\DashLine(95,115)(85,115){2}
\Line(85,115)(70,115)
\Line(70,115)(70,145)
\DashLine(85,115)(85,145){2}
\DashLine(85,145)(70,145){2}
\Line(70,145)(60,145)
\Line(60,145)(60,165)
\DashLine(70,145)(70,165){2}
\DashLine(70,165)(60,165){2}
\Line(60,165)(45,165)
\Line(45,165)(45,185)
\DashLine(60,165)(60,185){2}
\DashLine(60,185)(45,185){2}
\Line(45,185)(35,185)
\Line(35,185)(35,193)
\DashLine(45,185)(45,193){2}
\DashLine(45,193)(35,193){2}
\Line(35,193)(0,193)
\DashLine(35,193)(35,200){2}
\LongArrow(100,200)(100,100)
\LongArrow(100,200)(0,200)
\LongArrow(100,100)(0,100)
\LongArrow(100,100)(100,0)
\LongArrow(200,100)(100,100)
\LongArrow(200,100)(200,0)
\EBox(0,0)(100,200)
\EBox(0,0)(200,100)}
\end{picture} \caption{An abstract picture of the 3-parton geometry. The area below the zigzag line corresponds to the unfragmented string. The dashed lines indicate hadrons that are about to be formed.} \label{fig:abstract}
\end{center}
\end{figure}
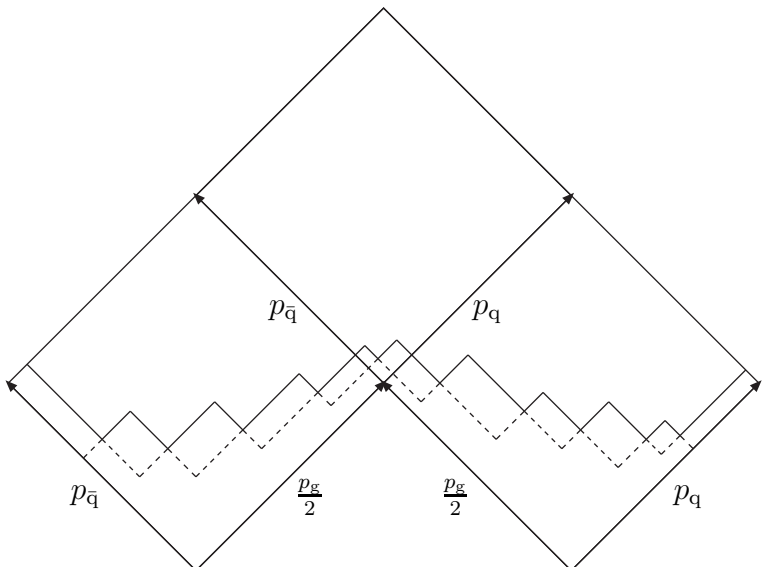 

When one becomes familiar with the geometry of Fig.~\ref{fig:diamond} one can proceed by flattening out the first three regions, as shown in Fig.~\ref{fig:abstract}, so that an abstract but clear-cut picture emerges. This has to be done by cutting along the gluon 4-vector; hence it is mapped onto two vectors in this diagram. Also the 4-vectors are rescaled and put in perpendicular directions, corresponding to boosts into local rest frames.  It is now easy to see the resemblance with the 2-parton case, and fragmentation is done as usual in each region, which is indicated in Fig.~\ref{fig:abstract}. The hadrons that cross over region boundaries are dealt with by simply adding the two separate region contributions; together the regions contribute to the full mass-square and 4-momentum of this hadron. A main complication is that the $z$ variable introduced in eq.~(\ref{zed}) has to be generalized to allow for steps not only inside but also between regions. This can be done by relating $z$ to the proper time of $\mathrm{q}_i\mathrm{\bar{q}}_i$~\cite{desy}, but we will not go into details here.

 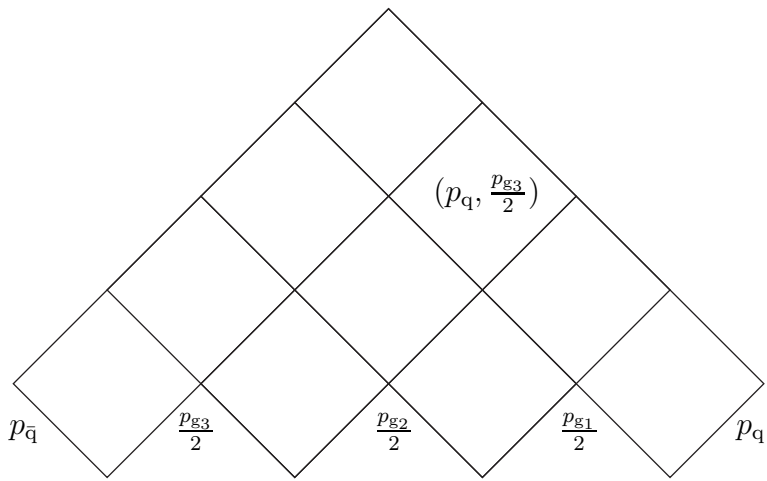
\begin{figure} \begin{center} \begin{picture}(200,200)(-100,-200)
\Text(80,-160)[]{$\frac{p_\mathrm{g_1}}{2}$}
\Text(145,-160)[]{$p_\mathrm{q}$}
\Text(10,-160)[]{$\frac{p_\mathrm{g_2}}{2}$}
\Text(-65,-160)[]{$\frac{p_\mathrm{g_3}}{2}$}
\Text(-130,-160)[]{$p_\mathrm{\bar{q}}$}
\Text(45,-70)[]{$(p_\mathrm{q},\frac{p_\mathrm{g_3}}{2})$}
\rotatebox{-135}{
\EBox(0,0)(200,50)
\EBox(0,50)(150,100)
\EBox(0,100)(100,150)
\EBox(0,0)(50,200)
\EBox(50,0)(100,150)
\EBox(100,0)(150,100)}
\end{picture} \caption{A $5$-parton `pyramid' that illustrates how an $n$-parton 
case might look in the abstract formalism. The four-vector pair that defines one 
region is shown as an example.} \label{fig:njet}
\end{center}
\end{figure}
 
It then becomes an easy task to generalize this discussion into an $n$-parton system, shown in Fig.~\ref{fig:njet} for $n=5$. Like the 3-parton case there will be more allowed regions than one might first expect. The number of regions equals the number of pairs of partons you can pick from the given $n$ particles, disregarding the order in which they are picked. Thus there are $n(n-1)/2$ available regions and they can be identified by the pair of parton 4-vectors that span them, i.e. $(p_i,p_j)$. Fragmentation can occur in all the regions, but is less likely the further up in the `pyramid' you are, and of course depends on whether the string has had time to fragment in lower regions, or did not have the time to do that. 
  
\subsection{The LUBOEI algorithm} \label{sec:luboei}

There are several ways to introduce Bose--Einstein effects into a hadronization model. The most direct approaches are to define some simplified complex amplitudes, from wave functions or matrix-elements, that can be symmetrized for identical bosons before calculating the needed probabilities. According to Quantum Mechanics, these probabilities are proportional to the square of the absolute value of the matrix element, times the appropriate phase space factors. These models are often said to be of the global type since each event is associated with a weight that depends on the production of all the particles in the event. Global models has been tried out by people in Lund~\cite{bo} and by others, but they all struggle with computational limitations and some inconsistencies that the models produce.

The models of local type alters the correlation of 4-momenta between identical pairs of mesons within an event, and in contrast to the global types all generated events have unchanged weight $=1$. The model worked out by Sj\"ostrand \& L\"onnblad~\cite{luboei}, sometimes called LUBOEI, is of this type. The local approach is more brute-force than the global one, but is computationally so much simpler that it has allowed more detailed studies. 
 
It has been found that a sensible way to account for Bose--Einstein effects is to introduce a two-particle\footnote{In the this section, whenever referring to `particle', we will mean boson, or, more specifically, we will always talk about pairs of identical mesons.} correlation function $c$ \cite{lorstad}. In its most simple form, which is useful for experimental purposes, it is a function of the four-momenta of the two particles. 
\begin{equation}
c(p_1,p_2)=\frac{\mathrm{P}(p_1,p_2)}{\mathrm{P}(p_1)\mathrm{P}(p_2)} = 
1+\lambda f(p_1,p_2) \label{corr}
\end{equation}
Here the numerator of the second expression is the double probability density to find particles with four-momenta $p_1$ and $p_2$, given that the combined wave function is correctly symmetrized. The denominator is the product of two independent one-particle probability densities. In the limit where there is no Bose--Einstein effect this ratio ought to be unity. However, correlations from the jet structure of the event, energy--momentum conservation, etc., can in reality alter this significantly. In practice one therefore is more likely to define $c(p_1,p_2)= P_\mathrm{BE}(p_1,p_2)/P_\mathrm{NoBE}(p_1,p_2)$, where the numerator corresponds to the real world, with Bose--Einstein effects, and the denominator to  a hypothesized one, identical in every respect except the need to perform a Bose--Einstein symmetization. A suitable parameterization is then the last expression in eq.~(\ref{corr}), where $\lambda$ is an \emph{ad hoc} parameter; naively $\lambda=1$ if $f$ is normalized to $f(p_1=p_2)=1$, but  screening effects, such as when the two particles originates from different sources, will reduce the Bose--Einstein effects and give an effective $\lambda < 1$. 

The function $f (p_1,p_2)$ contains all the physics and should mainly depend on the invariant four-difference, $Q=\sqrt{-(p_1-p_2)^2}$, of the momenta $p_1$ and $p_2$. Goldhaber et al. \cite{goldhaber} derived an approximate form, when considering a Gaussian shaped source, $\exp(-r^2/2R^2)$, of radius $R$: \begin{equation}
f(Q) \approx \mathrm{e}^{-Q^2R^2} \label{gauss}
\end{equation}
In our studies we alternatively allow a correlation function where $f(Q)$ is an exponential factor, instead of the Gaussian factor in eq.~(\ref{gauss}), and the final expression for $c(Q)$ is the one given by eq.~(\ref{first}).
 
The experimental value of  $R$ is about 1 to $0.5$ fm, or 5 to $2.5 \ \mathrm{GeV}^{-1}$;thus the region where the Bose--Einstein effects are present is very small, $Q< 0.2-0.4$ GeV, compared to the dimensions of an extended string ($\sim 91$ GeV at LEP1),  and is comparable to the width of the string (or for Q, the transverse momenta of quarks). 

As a starting point we now consider uncorrelated particle production, uniformly in phase space, $\propto \mathrm{d}^3 p_i/E_i$. A boost to the rest frame of a pair reveals that the relative momentum then is distributed according to
\begin{equation}
\frac{\mathrm{d}^3Q}{E_{\mathrm{pair}}} \propto \frac{Q^2 
\mathrm{d}Q}{\sqrt{4m^2+Q^2}} \label{density}
\end{equation} 
where $m$ is the mass of a single meson. When Bose--Einstein effects are included, the expression in eq.~(\ref{density}) has to be multiplied by the correlation function $c(Q)$ to give the right distribution. A pragmatic way to include Bose--Einstein effects, when given a $Q$-distribution without such correlations, is to perform a local shift $Q 
\rightarrow Q -\delta (Q)$  given by the equation
\begin{equation}
\int^Q_0 (1+\lambda\mathrm{e}^{-(qR)^{\eta}})\frac{q^2\mathrm{d}q}{\sqrt{4m^2 +q^2}}=\int^{Q+\delta(Q)}_0 \frac{q^2 \mathrm{d}q}{\sqrt{4m^2 +q^2}}
\end{equation}
This may be solved  approximately for small $\delta$'s, which yield
\begin{equation}
\delta (Q)=\lambda \frac{\sqrt{4m^2 +Q^2 }}{Q^2} \int^Q_0 \frac{q^2}{\sqrt{4m^2 
+q^2 }} \mathrm{e}^{-(qR)^{\eta}}\mathrm{d}q
\end{equation}

Note that the algorithm does not change the number of particles, which implies that the effective $c(Q)$ is less than unity for intermediate $Q$, contrary to the parameterization of eq.~(\ref{corr},\ref{gauss}). In the LUBOEI algorithm a shuffling procedure transforms the pairwise relative shift into a total shift for each hadron, owing to the net effect of all the pairs the hadron belongs to, such that the total momentum of the event is conserved. This also introduces some non-linear effects in the model, e.g. non-trivial three-particle correlations. Unfortunately, these types of operations cannot be done without violating either the conservation of energy or the conservation momentum of the total system.  This is perhaps the biggest disadvantage with local models. The problem is dealt with by scaling all the hadron 3-vectors with a common factor in the rest frame of the event. However, there are other more sophisticated ways to shuffle the hadron 4-momentum, such as the one described in~\cite{luboei}, which could be argued to be more realistic, but for our purposes this simple procedure is enough.

\section{Identifying $R$ with a space--time scalar $S$}        

In the Goldhaber derivation of the correlation function, $R$ is the radius of a spherical symmetric Gaussian source out of which pions emerge. At first glance the Lund String Model seems to lack such a source; instead the string is more of a cigar-shaped source, thus making it difficult to motivate why this correlation function should work well for 2-jets. It has been shown that the LUBOEI algorithm can reproduce experimental data with $\lambda=1.35$ and $R=0.6$ fm~\cite{delphi}, which tells us three things: First, the model is not flawless, since $\lambda$ should be at most unity in the interpretation given in section~\ref{sec:luboei}; thus it is compensating for some physics that is still unaccounted for. Second, $R$ is approximately the separation between neighbouring (neutral) and next to neighbouring (neutral \& charged) pions at time of creation in the 2-parton picture. This is an interesting fact since pions are the hadrons with the highest multiplicities and thus give the dominating contribution to the Bose--Einstein effects. Third, $R$ is comparable with width of the string ($\sim 0.7$ fm) within errors; hence if $R$ is to be interpreted as an effective source radius then this source is small enough to fit inside the apparent cigar-shaped source, which makes is possible to consider that it has a spherical symmetric distribution as a first approximation. The are however experimental data that indicate that the source has a certain elongation in the sting direction~\cite{amsterdam} as one might suspect, but we choose to ignore this small correction in this work. It should also be noted that physical considerations require that the Bose--Einstein correlation between well-separated hadrons is unmeasurably small, whichever source form is used; this explains the apparent misfit of the effective Bose--Einstein source and the cigar-shaped string region.    

If the effective source is not equal to the apparent source, defined by the string geometry, then it is no longer necessary that the size of the effective source is constant for different pairs of hadrons within an event. The problem that Bose--Einstein effects for higher multiplicities of jets are not well described by the current LUBOEI algorithm indicate that the spatial structure of the string is important and that information about this should be invoked in the correlation function. The simplest way to achieve this is to parameterize $c$ with two independent variables $Q$ (kinetic) and $S$ (spatial), where $Q$ is the invariant 4-momentum difference as before and $S$, replacing $R$, could be something like the invariant 4-position difference of the hadron production vertices, $S=\sqrt{-(v_2-v_1)^2}$, see further discussion below. Keeping an expression very similar to the $Q$-parameterization of the correlation function, $c(Q,S)$ becomes
\begin{equation}
c(Q,S)=1+\lambda e^{-(kQS)^\eta} \label{cspatial}
\end{equation} 
where $k$ is a dimensionless constant, which is to be fitted to data.

\subsection{Production vertices of hadrons} \label{sec:prod}

The experimental accuracy allows only a certain crude classification of production vertices (the points from which the particles seem to originate). Most particles appear to come from a common primary vertex, and only long-lived particles that have time to travel at least some millimetres give rise to resolved secondary vertices when they decay. Examples of the latter are D and B mesons, $\mathrm{K^0_s}$ and $\Lambda^0$, and particles like $\mathrm{K}^\pm$ and $\pi^\pm$ that decay outside the detector. 

For theoretical purposes the experimental primary vertex is large enough to be resolved into separate productions regions for primary hadrons (hadrons that directly emerge out of the string fragmentation process) and secondary hadrons emerging from fast decays of the primary hadrons, e.g. $\rho$, $\mathrm{K}^*$, $\eta$ and higher excitations. The Heisenberg uncertainty principle prevents us from talking about a specific point where a hadron is produced with certainty, but it still allows us to speak about the mean value of a production point as if it was a classical quantity. Thus we can use the convention that a production vertex of a hadron is to be taken as the average of the area where the probability to produce the hadron in question is non-vanishing. 

\begin{figure}
\begin{center}
\begin{picture}(200,180)(-100,10)
\rotatebox{45}{
\SetScale{1.5}
\Line(0,0)(0,75)
\Line(0,0)(90,0)
\Line(59,12)(100,12)
\DashLine(95,4)(59,4){2}
\DashLine(59,4)(59,12){2}
\Line(29,36)(59,36)\Line(59,12)(59,36)
\DashLine(59,12)(29,12){2}
\DashLine(29,12)(29,36){2}
\Line(29,12)(59,36)
\GBox(59,36)(89,60){0.8}
\GBox(89,60)(119,84){0.8}
\Vertex(29,12){2}
\Vertex(59,36){2}
\Vertex(44,24){2}
\Line(19,71)(29,71)\Line(29,35)(29,71)
\DashLine(29,36)(19,36){2}
\DashLine(19,36)(19,71){2}
\Line(19,71)(19,92)
\DashLine(19,71)(4,71){2}
\DashLine(4,71)(4,80){2}
\SetScale{1}}
\Text(28,128)[]{\scriptsize hadron}
\Text(-5,95)[cb]{\scriptsize production}
\Text(-5,95)[ct]{\scriptsize region}
\Text(0,30)[]{\scriptsize string}
\Line(-5,86)(8,65)
\Text(30,43)[]{2}
\Text(33,72)[]{3}
\Text(36,100)[]{1}
\end{picture}
\end{center} \caption{The three candidates for a hadron production vertex. This picture represent the same 2-parton case as shown in Fig.~\ref{fig:2jet}} 
\label{fig:3vertex}
\end{figure}
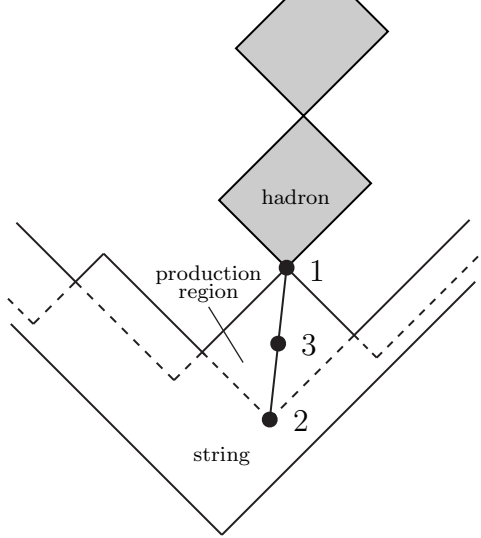

Given that this area is approximately the same as the one that appears in the dashed region of Fig.~\ref{fig:3vertex}, there are three natural points associated with a hadron production vertex. An obvious choice, no. 1, is the point where the two constituent quarks meet for the first time. A second natural choice, no. 2, is the latest point which is casually connected to both surrounding $\mathrm{q\bar{q}}$-vertices. The third point, no. 3, is simply the average between the first and second point, which is the same as the average between the two $\mathrm{q\bar{q}}$-vertices, i.e. $v_i=(w_{i}+w_{i-1})/2$, where $w_i$ for $\mathrm{q}_i\mathrm{\bar{q}}_i$ production and $v_i$ for hadron $i$. By symmetry the production vertex should lie somewhere on the line that connects the first and the second point, and there is no reason to believe that a hadron is produced earlier than the second point or later than the first point. If the production vertex should be considered to be an average of the possible production points then the third point is in some favour. The quantum mechanical considerations in \cite{bo} also favour this choice.

\begin{figure}
\begin{center}
\epsfig{file=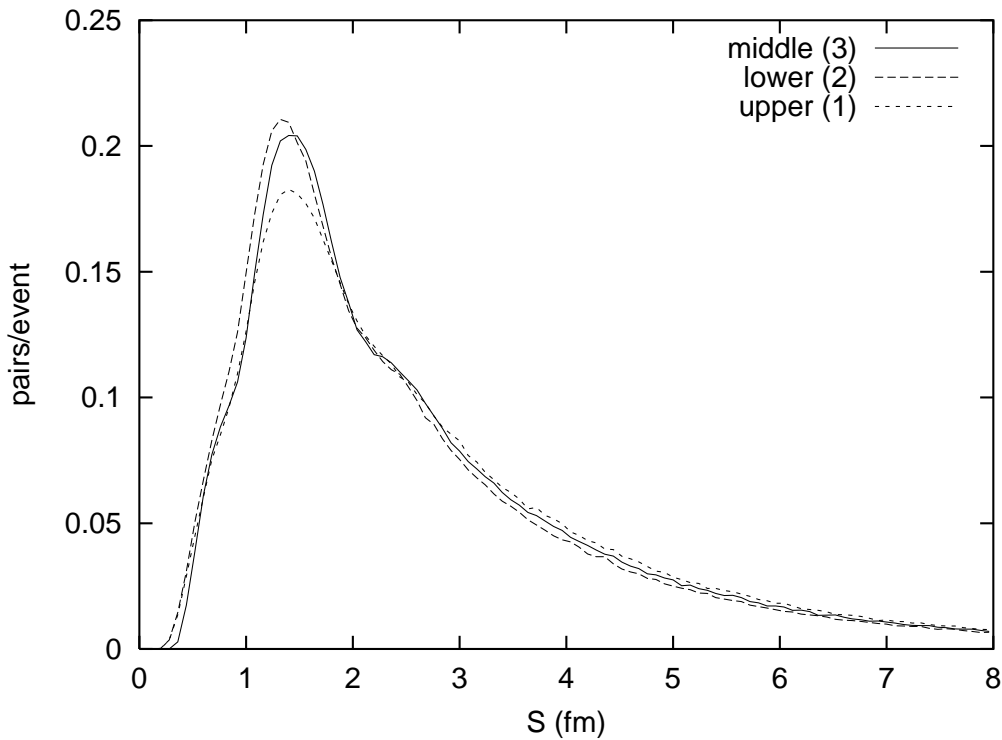, height=100mm}
\caption{Distribution of $S=\sqrt{-(v_2-v_1)^2}$ for pairs of charged identical primary pions with $Q<0.5$ GeV for a 2-parton event (only u-quarks, $\mathrm{u\bar{u}}$) at $E_\mathrm{cm}=91$ Gev. The three different cases correspond to the possible hadron production vertices shown in Fig~\ref{fig:3vertex}} \label{fig:plot1} 
\end{center}
\end{figure}

\begin{figure}
\begin{center}
\epsfig{file=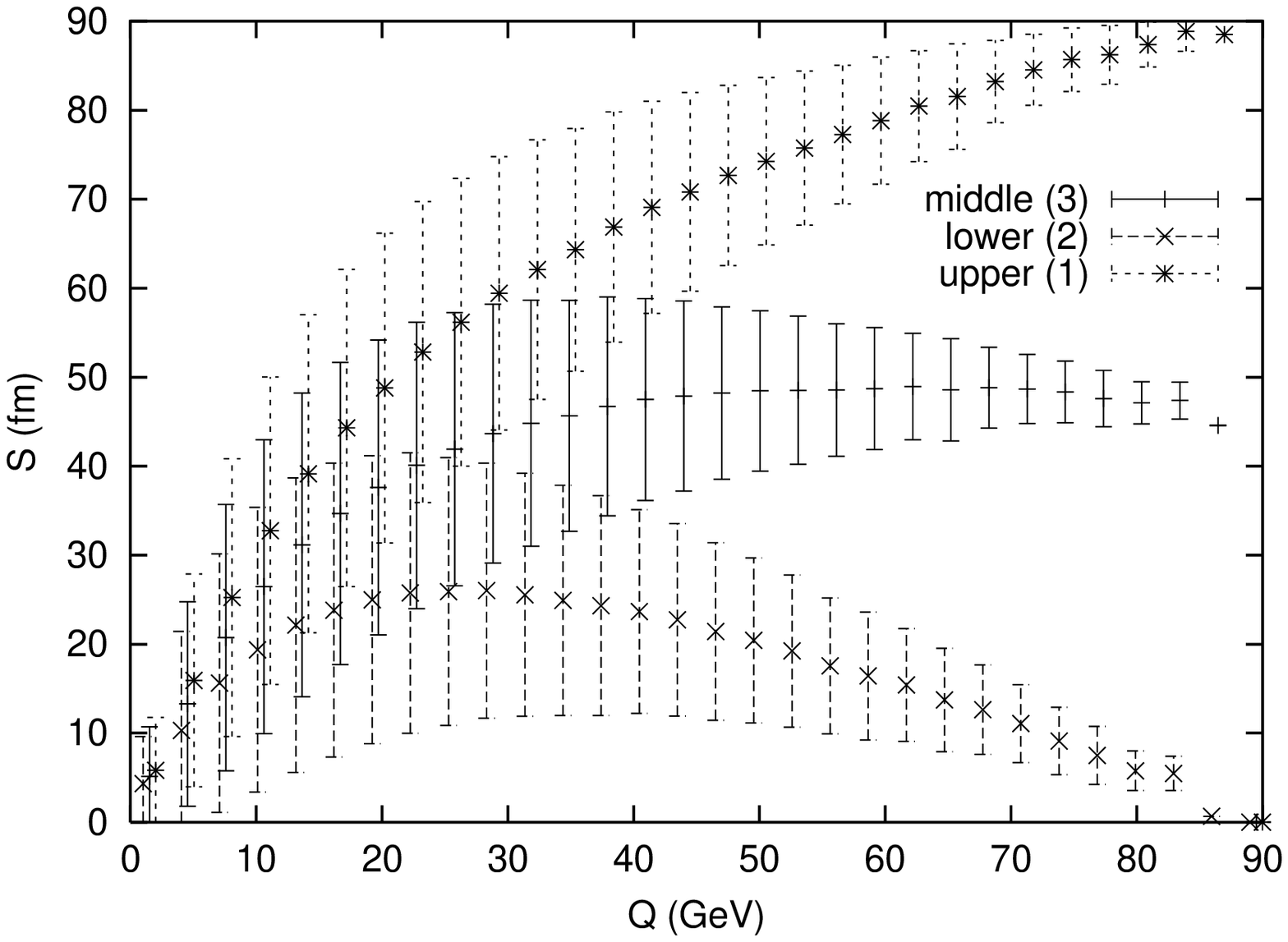, height=100mm}
\caption{The mean $\langle S \rangle$ shows that there is an approximate correspondence between $Q$ and $S$ for charged identical pions. Here for the 2-parton case ($E_\mathrm{cm}=91$ GeV, $\mathrm{u\bar{u}}$) and for three different definitions of the hadron production vertex. The error-bars give the one-sigma spread in each bin.} \label{fig:plot2} 
\end{center}
\end{figure}

Fig.~\ref{fig:plot1} is produced using \textsc{Pythia}~\cite{pythia}, the Lund Model event generator, and shows that for low-$Q$ pairs there is essentially no difference between $S$ distributions of the three possible vertex definitions. It is mainly for pairs with low $Q$ that the Bose--Einstein correlation is important according to eq.~(\ref{cspatial}), thus the physics will not crucially depend upon which choice we make. As a curiosity Fig.~\ref{fig:plot2} shows the average of $S$ as a function of $Q$ in the fully allowed range. For $Q$ larger than $\sim E_\mathrm{cm}/10$ there is a clear divergence in the three vertex definitions, and at the extreme $Q= E_\mathrm{cm}$ the different $S$ are completely fixed with the values $E_\mathrm{cm},0 , E_\mathrm{cm}/2$ for points 1 to 3 respectively.  

From a theoretical point of view the third point seems more appealing, but the choice is not crucial for our discussion. As default we will chose the third point when talking about the associated production vertex for primary hadrons in the following pages.

The fast decays of resonances, i.e. of $\rho$ and $\mathrm{K}^*$ in practice, 
might also contribute to Bose--Einstein effects, hence production vertices for 
the decay product must be deduced from simple principles. There are two ways of 
doing this:

\begin{enumerate}
\item{It can be argued that most of these decays are fast enough so that it is a good approximation to assume that the resonance decays immediately; in this sense the decay products are part of the primary hadrons and the resonance is merely an enhancement of the pair/triplet-mass distribution. The calculation of production vertices is then easily taken care of by inserting extra $\mathrm{q\bar{q}}$-vertices where appropriate inside the resonance production region, such that kinematical quantities are preserved.}

\item{A more realistic picture would be obtained if the resonances where allowed to live for a proper time $\tau$, which is an exponentially distributed random variable with expectation value $\tau_0=1/\Gamma$, $\tau_0$ being the lifetime of a particle and $\Gamma$ its width. The decay vertex of a resonance, which in this simplified treatment could be taken as the production vertex of the daughter particles, is then given by:
\begin{equation}
v_\mathrm{decay}=\frac{p_\mathrm{r}\tau}{m_\mathrm{r} \kappa}+v_\mathrm{r}
\end{equation}
Here r stands for resonance and $v_\mathrm{r}$ is the production vertex of the resonance.}
\end{enumerate}
The production point assignment of a secondary hadron could be taken as an average of the possible decay states. For a decay resulting in two particles the two decay states (in 1+1 dimensional space--time) are each others space reflections in the rest frame of the decay; thus the average production point must lie on the extrapolated world line of the decaying particle. The decay point, with method no. 2, is then a good approximation of the production point for both daughter particles. A similar argument hold in real space--time and for more than two decay products. This assignment would be unfortunate if any two of the decay products where identical mesons, since it would result in an overestimate of the Bose--Einstein correlation. As it turns out, none of the fast decaying resonances have allowed decays that would give this problem.

\subsection{A Lorentz invariant similar to the production vertex difference}

When dealing with decays according to the second method (used as default) one runs into problems, since hadron vertices are no longer separated with strictly space-like distances. Quite often decay vertices end up inside the future light-cone of one or several of the primary hadrons. The ordinary invariant distance between production vertex ($S_\mathrm{ordinary}=\sqrt{-(v_2-v_1)^2}$) cannot be the correct spatial scalar to use in eq.~(\ref{cspatial}); firstly, $S$ is badly defined for time-like distances, and secondly, the Gaussian form of the correlation function will blow up for large time-like distances. All this could be avoided if the first method of assigning production vertices for decay products is invoked instead of the second method. But that would be to avoid the problem by choosing a worse description of the physical process. Furthermore, in multiparton configurations, it can even happen that a primary hadron lies inside the future light cone of another primary hadron from a different string region. 

The conclusion drawn is that we have to define a new effective spatial invariant $S_\mathrm{eff}$ that can be looked upon as the ordinary invariant 4-distance between two new effective vertices $v'_1$ and $v'_2$, which in turn are functions of the production vertices as well as other relevant parameters. The two requirement that must be fulfilled is: Firstly, $S_\mathrm{eff}$ and $S_\mathrm{ordinary}$ must be approximately equal for pairs of primary hadrons originating from the same string region, at least when the exponent of the correlation function is small, $kQS \leq 1$. Secondly, $S_\mathrm{eff}$ must necessarily be space-like, i.e. $v'_1$ and $v'_2$ must be simultaneous events in some reference frame.

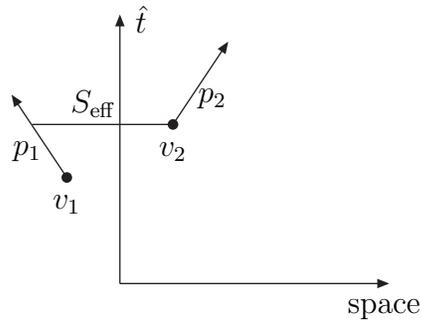
\begin{figure}
\begin{center}
\begin{picture}(200,105)(-100,0)
\Vertex(-20,40){2}
\Vertex(20,60){2}
\LongArrow(-20,40)(-40,70)
\LongArrow(20,60)(40,90)
\LongArrow(0,0)(0,100)
\LongArrow(0,0)(100,0)
\Line(20,60)(-33,60)
\Text(-35,50)[]{$p_1$}
\Text(35,70)[]{$p_2$}
\Text(-20,30)[]{$v_1$}
\Text(20,50)[]{$v_2$}
\Text(8,100)[]{$\hat{t}$}
\Text(100,-10)[]{space}
\Text(-10,68)[]{$S_\mathrm{eff}$}
\end{picture}
\end{center} \caption{A space-like scalar $S_\mathrm{eff} \sim S_\mathrm{ordinary} =-\sqrt{(v_2-v_1)^2}$} \label{fig:scalar}
\end{figure}

We have found it sensible to boost a pair of identical particles to their common rest frame and there define an effective Bose--Einstein distance $S_\mathrm{eff}$ to be the invariant spatial distance between the particles at the time of creation of the last particle. This distance, illustrated in Fig~\ref{fig:scalar}, then becomes
\begin{equation}
S_\mathrm{eff}^2=-(v_2-v_1-\frac{(v_2-v_1)\cdot(p_1+p_2)}{(p_1+p_2)\cdot p_1} 
p_1)^2 \label{invariant}
\end{equation}
where 1 is the early and 2 is the late particle. Here our choice of effective vertices were $v'_1=v_1+((v_2-v_1)\cdot(p_1+p_2)) p_1/((p_1+p_2)\cdot p_1)$ ($= v_1 + (t_2-t_1)p_1/E_1$ in pair rest frame) and $v'_2=v_2$. The new vertex $v'_1$ is then the point where particle 1 is found as particle 2 is simultaneously created in their common rest frame. It is clear from the definition that $S_\mathrm{eff}$ is always space-like.

\begin{figure}
\begin{center}
\epsfig{file=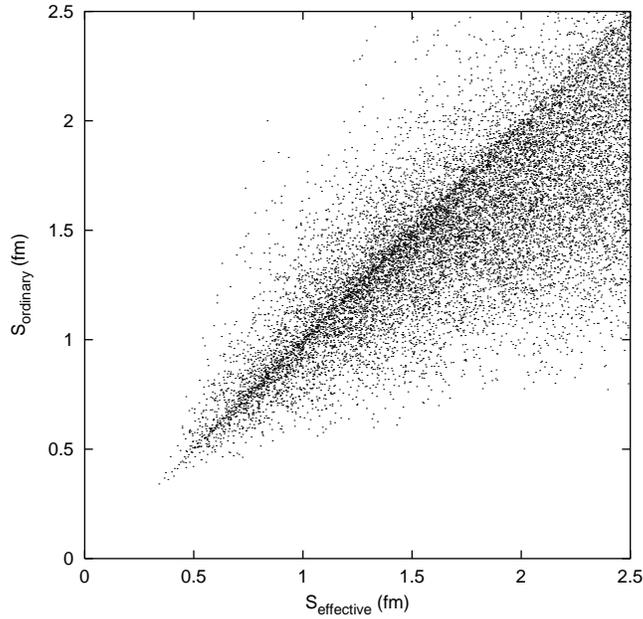, height=80mm}
\caption{The approximate correspondence between $S_\mathrm{ordinary}$ and $S_\mathrm{eff}$ for identical primary charged pions in the 2-parton case ($\mathrm{u\bar{u}}$).} \label{fig:scatter}
\end{center}
\end{figure}

Fig.~\ref{fig:scatter} shows a scatter plot comparing the ordinary invariant distance $S_\mathrm{ordinary}$ with the effective distance $S_\mathrm{eff}$ for primary charged pion-pairs, for the 2-parton case and in the region of interest. The correspondence is satisfactory; a large portion of the points fall on the ideal diagonal, and the plot is symmetric for $S_\mathrm{ordinary/eff}<1.5$ fm, which is appealing because on average no bias will show up. At large distances the data shows (region not fully included in plot) that $ S_\mathrm{eff}\geq S_\mathrm{ordinary}$ with increasing likelihood. 

From here on we shall only make use of the effective invariant distance defined in eq.~(\ref{invariant}) and hence we skip the index and refer to it simply as $S$.

\subsection{An algorithm to deduce primary production vertices}

Since the \textsc{Pythia} event generator does not keep in memory the primary production vertices, these have to be reconstructed from the output available. In this section we will show that the hadron 4-momentum, the rank of the hadrons and the original parton 4-momenta will give us all the information required to calculate the needed production vertices. The parton 4-momenta will give us information about the string geometry; each parton pair spans a sub-string (or sub-sheet). In the Lund Model the notion of rank is used to specify the space/flavour-order in which primary hadrons are produced. 

\subsubsection{The 2-parton algorithm}

In the two-parton case the algorithm is simple: enumerate the primary hadrons from right to left with a rank $i$ that runs from 1 to $m$. Call the $\mathrm{q\bar{q}}$-vertices $w_i$, where $w_0=k_0/\kappa$ is the turning point of the original quark, and $k_0$ is the 4-momentum of the same, as in section~\ref{sec:2part}. Then the first production vertex $v_1$, as defined by point no. 3 in Fig.~\ref{fig:3vertex}, is given by $v_1=(w_0+w_1)/2 = w_0+\tilde{p}_1/2\kappa$, where $\tilde{p}_1$ is defined by the relation $w_1= w_0+\tilde{p}_1/\kappa$. This can be generalized to any hadron production vertex $v_i$ with the recursive formulae:
\begin{eqnarray} \label{rec}
w_{i} & = & w_{i-1}+\frac{\tilde{p}_i}{\kappa} \nonumber \\
w_0 & = & \frac{k_0}{\kappa} \\
v_i & = & \frac{w_{i}+w_{i-1}}{2} \nonumber \\
i & = & 1 \ldots m \nonumber
\end{eqnarray} 
The translation vector $\tilde{p}_i$ between two neighbouring $\mathrm{q\bar{q}}$-vertices is the space-like diagonal of hadron $i$'s production region, and is related to the hadron 4-momentum $p_i$ in the following way:
\begin{equation} 
\tilde{p}_i= \frac{p_i \cdot k_0}{k_0 \cdot k'_0} k'_0- \frac{p_i \cdot 
k'_0}{k_0 \cdot k'_0} k_0
\end{equation}
where $k_0$ and $k'_0$ are the 4-momenta of the original quark and antiquark, respectively. We have in this equation assumed that $k_0$ and $k'_0$ are light-like vectors, i.e. the quarks are massless.  It can be shown, from conservation of the total system 4-momentum, that the recursion formulae give $w_m=k'_0/\kappa$, which is the correct answer. 

\subsubsection{The $n$-parton algorithm}

For the $n$-parton case the algorithm becomes much more technical, but the idea is the same as above. The physics discussion in subsequent sections can therefore be understood even if you choose to skip the rest of this section.

The procedure of the 2-parton case must now be performed in separate regions spanned by two partons, but it cannot be done independently since hadrons may cross region boundaries. Although there are $n(n-1)/2$ possible regions, the string fragmentation breakups can only occur in less than $2n-3$ of those regions, different for different events. This is due to the fact that fragmentation in an early region excludes fragmentation in some later regions, which are at time-like separations with respect to the first one. Hence a major complication, which one has to deal with in an $n$-parton algorithm, is to deduce in which regions the fragmentation ($\mathrm{q\bar{q}}$-production) actually occurred. To do this we need some definitions and conventions that are far from obvious.

With the risk of confusing\footnote{The notation in this section should be looked upon as being independent of the one used in section~\ref{sec:2part} although some parts are reused} the reader we define the $n$-parton `basis'-vectors to be $k_i$, $i=1\ldots n$, where $k_1$ is now the original quark 4-momentum and $k_n$ is the antiquark 4-momentum, but where the intermediary $k$'s are only half of the gluon momenta; and $m$ is no longer the total number of hadrons but the rank of a single hadron. 

If we take all the hadrons to be mesons, then a hadron $m$ is made up by a quark and an antiquark from two neighbouring breakups, which occur in regions $(k_{j'},k_{l'})$ and $(k_{j''},k_{l''})$ respectively.
We then use the convention that the expression 'hadron $m$ belongs to $(k_j,k_l)$' means that the relations $j' \le j \le j''$ and $l' \le l \le l''$ are fulfilled. An alternative interpretation of this expression is that the 4-momentum of hadron $m$ has a component that can be written as a linear combination of $k_j$ and $k_l$. With this convention the region with largest indices, $j$ and $l$, that a hadron can belong to will be the region where the production of the antiquark occurred, and the region with the lowest indices will be the one where the production of the quark occurred. 

Then we need to define a scalar indicator $z$, that gives the fraction of the basis-vector $k_j$ (or $k_l$ if $j$ and $l$ are interchanged) that is used by the $m$ first hadrons. In order to make this quantity well defined one has to assume that we already know that hadron $m$ belongs to $(k_j,k_l)$. [From the onset we know this to hold for the region of the antiquark vertex of hadron $m-1$.] Then $z$ is given by:
\begin{equation}
z(j,l,m)=(\sum_{i=1}^m p_i-\sum_{i=1}^{j} k_i-\sum_{i=1}^{l} k_i+k_1+k_j 
+k_l) \cdot  \frac{ k_l}{k_j \cdot k_l}
\end{equation}

From the definition it might not be clear what the purpose of this indicator is, but it is intended to be used in the following way: if
\begin{equation}
\begin{array}{ccl}
z(j,l,m)>0 \mbox{ and } z(l,j,m)<1 & \Longrightarrow & \mbox{hadron } m \mbox{ belongs to } (k_j,k_l) \\
z(j,l,m)<0 \mbox{ and } z(l,j,m)<1 & \Longrightarrow & \mbox{hadron } m \mbox{ belongs to } (k_{j+1},k_l) \\
z(j,l,m)>0 \mbox{ and } z(l,j,m)>1 & \Longrightarrow & \mbox{hadron } m \mbox{ belongs to } (k_j,k_{l+1}) \\
z(j,l,m)<0 \mbox{ and } z(l,j,m)>1 & \Longrightarrow & \mbox{hadron } m \mbox{ belongs to } (k_{j+1},k_{l+1})
\end{array} \label{criteria}
\end{equation}
In either of the three last cases, the procedure is to be iterated from the new region defined on the right hand side. If the indicator gives the first case one knows that the antiquark of hadron $m$ was produced in the region defined by this case; thus implying that hadron $m+1$ belongs to this region, and the procedure can be continued with this hadron. We have here assumed that $1\le j<l\le n$ (in the definition of $z$ this is not assumed) since this will give us the correct number of regions. This implies that some of the regions on the right hand side will not exist, but with the physical picture described so far the left hand side criteria will never be fulfilled unless the right hand side statements make sense. Thus with the $z$-indicator it is possible for every hadron to iteratively deduce which regions it belongs to, which in turn will give which region a specific $\mathrm{q\bar{q}}$-vertex ($w_m$) belongs to.
 
With the definition of rank the first hadron will always belong to the first region (right-most region in Fig.~\ref{fig:njet}) labelled by $(k_1,k_2)$. The vertices are then calculated with formulae that can be made recursive like to those of eq.~(\ref{rec}), but are better displayed in a direct form: 
\begin{eqnarray} \label{rec2}
w_m & = & z(l,j,m) \frac{k_l}{\kappa}-z(j,l,m) \frac{k_j}{\kappa} + \sum_{i=j}^{l-1}\frac{k_i}{\kappa}\nonumber \\
w_0 & = & \frac{k_1}{\kappa} \\
v_m & = & \frac{w_{m}+ w_{m-1}}{2} \nonumber\\
m & = & 1 \ldots \sharp \mbox{ primary hadrons} \nonumber
\end{eqnarray}
Here $j<l$ are chosen so that $w_m$, which is the antiquark production vertex of hadron $m$, belongs to region $(k_j,k_l)$; thus $j$ and $l$ are functions of $m$, determined with the aid of the indicator $z$. As before $v_m$ is the $m$'th hadron production vertex. 

\begin{figure}
\begin{center}
\epsfig{file=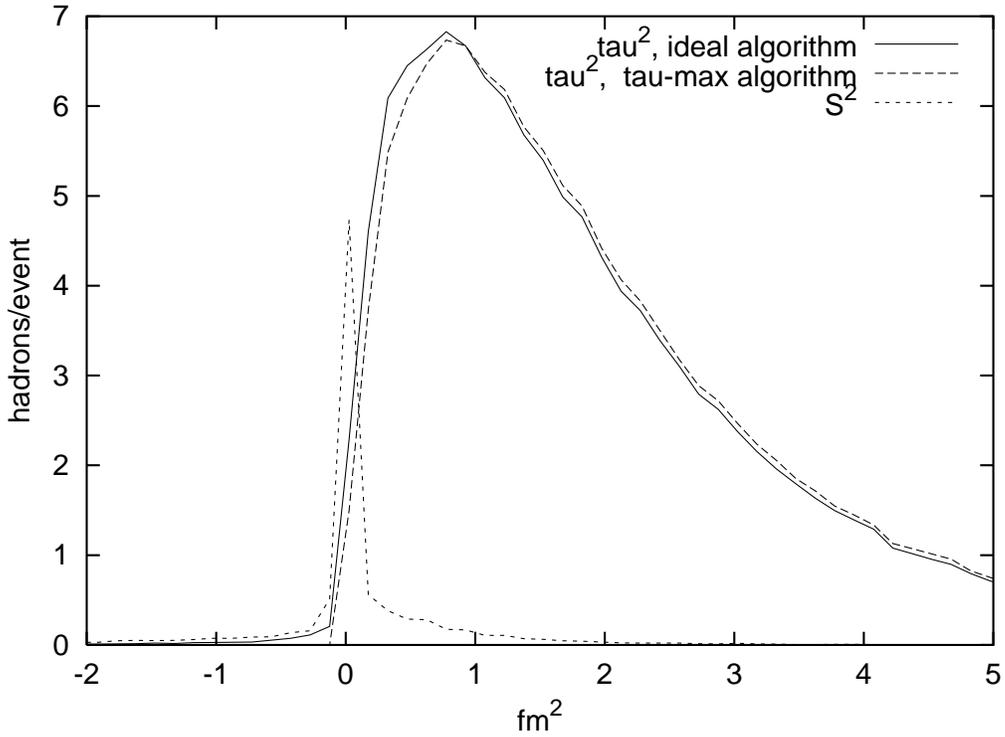, height=100mm}
\caption{The $\tau^2$-distribution (where $\tau$ is the proper time of hadron production vertex) for the ideal algorithm and the $\tau$-maximizing algorithm in a 3-parton `Mercedes' event with $E_\mathrm{cm}=91$ GeV ($\mathrm{u\bar{u}}$).  The distribution $S^2$ here refers to the usual invariant distance between the two production vertices calculated by the two algorithms for every primary hadron. Note 1: a delta function contribution at the origin, corresponding to identical vertices, is not shown in the $S^2$ curve. Note 2: The apparent negative $\tau^2$'s in the $\tau$-max algorithm is an effect caused by the drawing routine and the choice of bin-size, not a real result.} \label{fig:3jettau}
\end{center} 
\end{figure}
\begin{figure}
\begin{center}
\epsfig{file=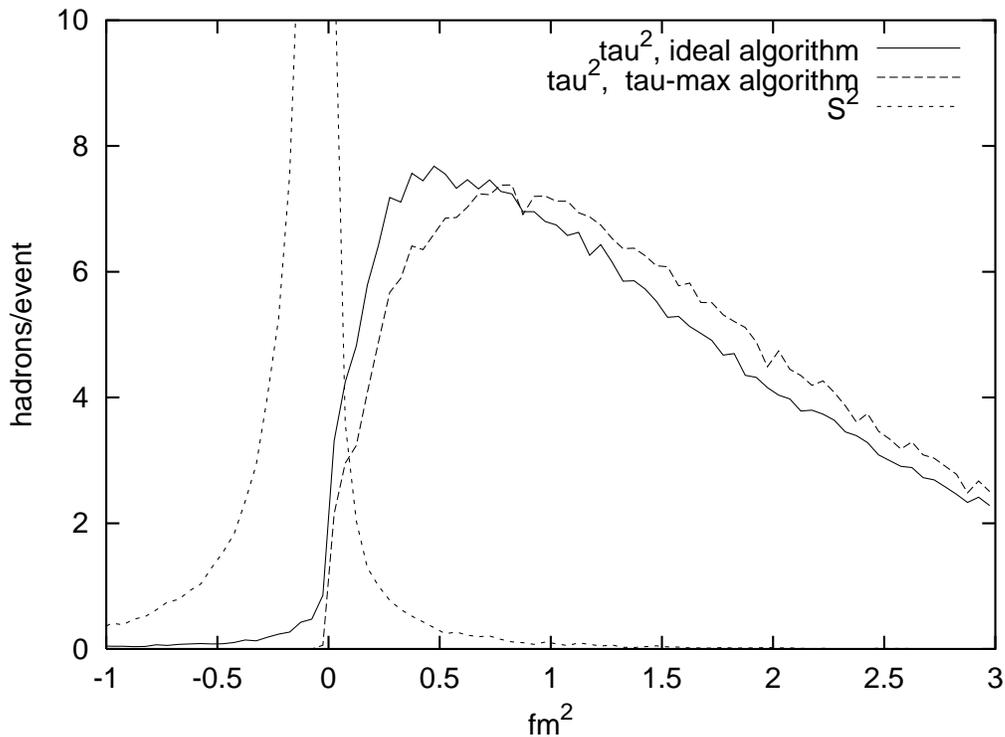, height=100mm}
\caption{As Fig.~\ref{fig:3jettau}, but with a physical mixture of $n$-parton events at $E_\mathrm{cm}=91$ GeV (using PYEEVT, only $\mathrm{u\bar{u}}$).} 
\label{fig:njettau2} 
\end{center}
\end{figure}
With an ideal $n$-parton system, such as described in section~\ref{sec:3jet}, the vertex algorithm should be perfectly flawless. In practice, when using an event generator many further physical aspects are included, that might or might not be important for our study, but cannot simply be turned off and will inevitably give rise to complications. It was discovered when using \textsc{Pythia} that for $n>2$ our algorithm, with the strict criteria in eq.~(\ref{criteria}), could not always choose the correct region to place a hadron in. A small fraction of the hadrons were assigned production vertices with a negative proper time, which is not acceptable. This can be related to ambiguities in the hadronization algorithm itself~\cite{desy}, where e.g. transverse momentum fluctuations can lead to contradictory assignments of production region. A pragmatic solution to this problem was to evaluate for each hadron several possible production vertices, corresponding to every possible region spanned by two partons, and then to pick the vertex with largest proper time. In principle you evaluate the $w_m$ and $v_m$ for every region $(k_j,k_l)$ with the formulae in eq.~(\ref{rec2}), and then compare for each $m$ all the $\tau^2=v_m^2$ given by different $j$ and $l$. A further minor approximation, to reduce the computing time as well as the code, resulted in that only the two lowest levels of regions $(1 \leq l-j \leq 2)$ were evaluated in the process. The result was not a bias for high proper times as one might expect; instead the modified algorithm and the ideal algorithm produced in the 3-parton case a very similar distribution of the proper time, shown in Fig.~\ref{fig:3jettau}, with the exception that the former was liberated from a tail of negative values. For a physical mixture of $n$-parton events, there is a small but clear discrepancy between the definitions for the majority of data points, shown in Fig.~\ref{fig:njettau2}. This is caused by the large number of soft gluons, which make the space--time picture fuzzy, thus undermining the validity of the ideal algorithm. Nevertheless, the two algorithms can be considered as being equally good (or bad) in determining the correct vertices, and the discrepancy can be interpreted as a measure of the uncertainty in the physical understanding. Henceforth, we choose to use only the modified algorithm and avoid any negative proper times. Note, even though we have assumed all quarks to be massless in the recipe of this algorithm as presented here, in the algorithm actually used to produce the results we also accounted for heavy quarks by finding the light-like vectors that span a given parton region, with $\tau=0$ at the primary production vertex. 

\subsection{The shift $\delta$ as a two-parameter function}

\begin{figure}
\hspace*{-0.5cm}
\epsfig{file=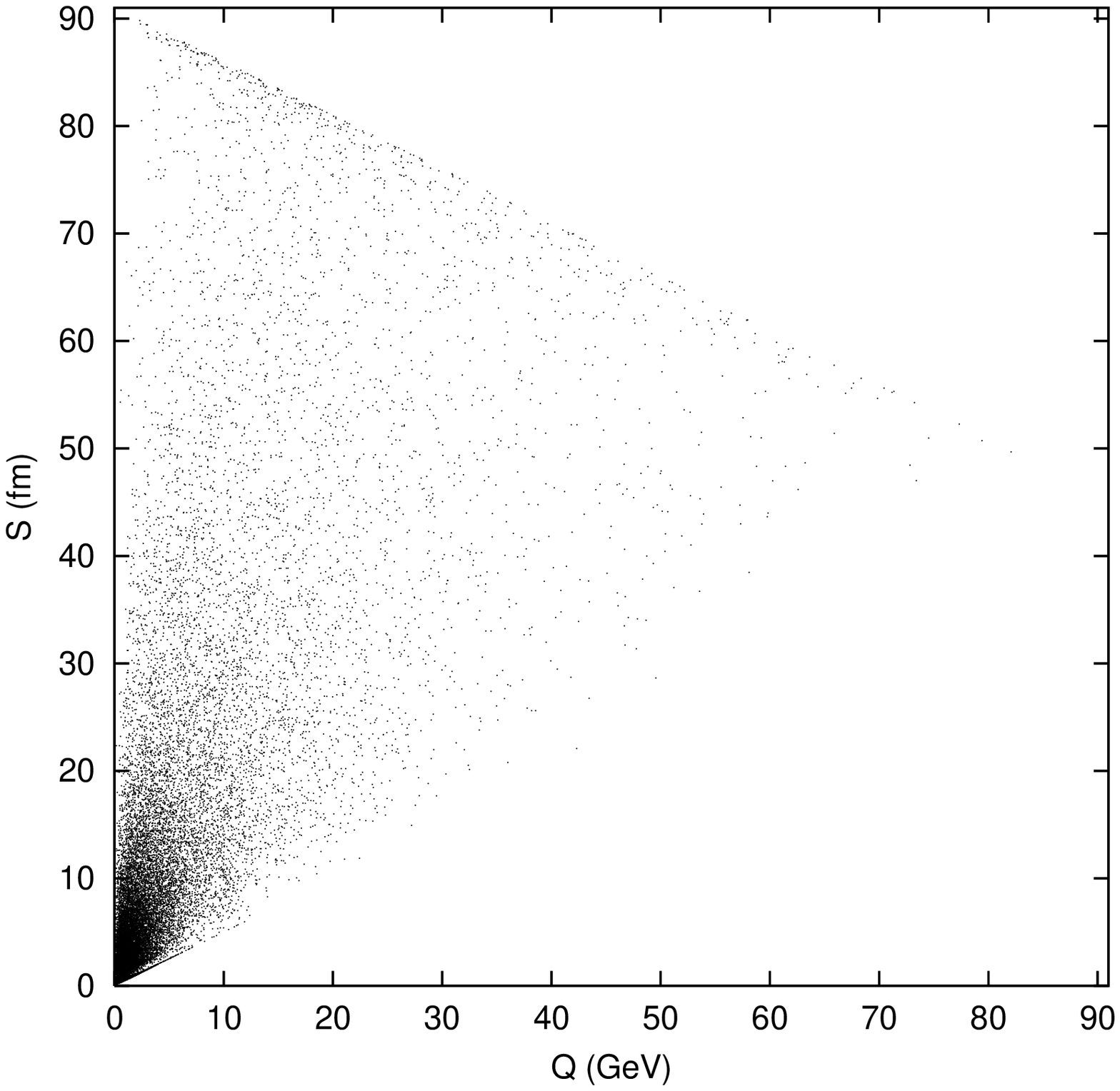, height=80mm}
\hspace*{-0.5cm}
\epsfig{file=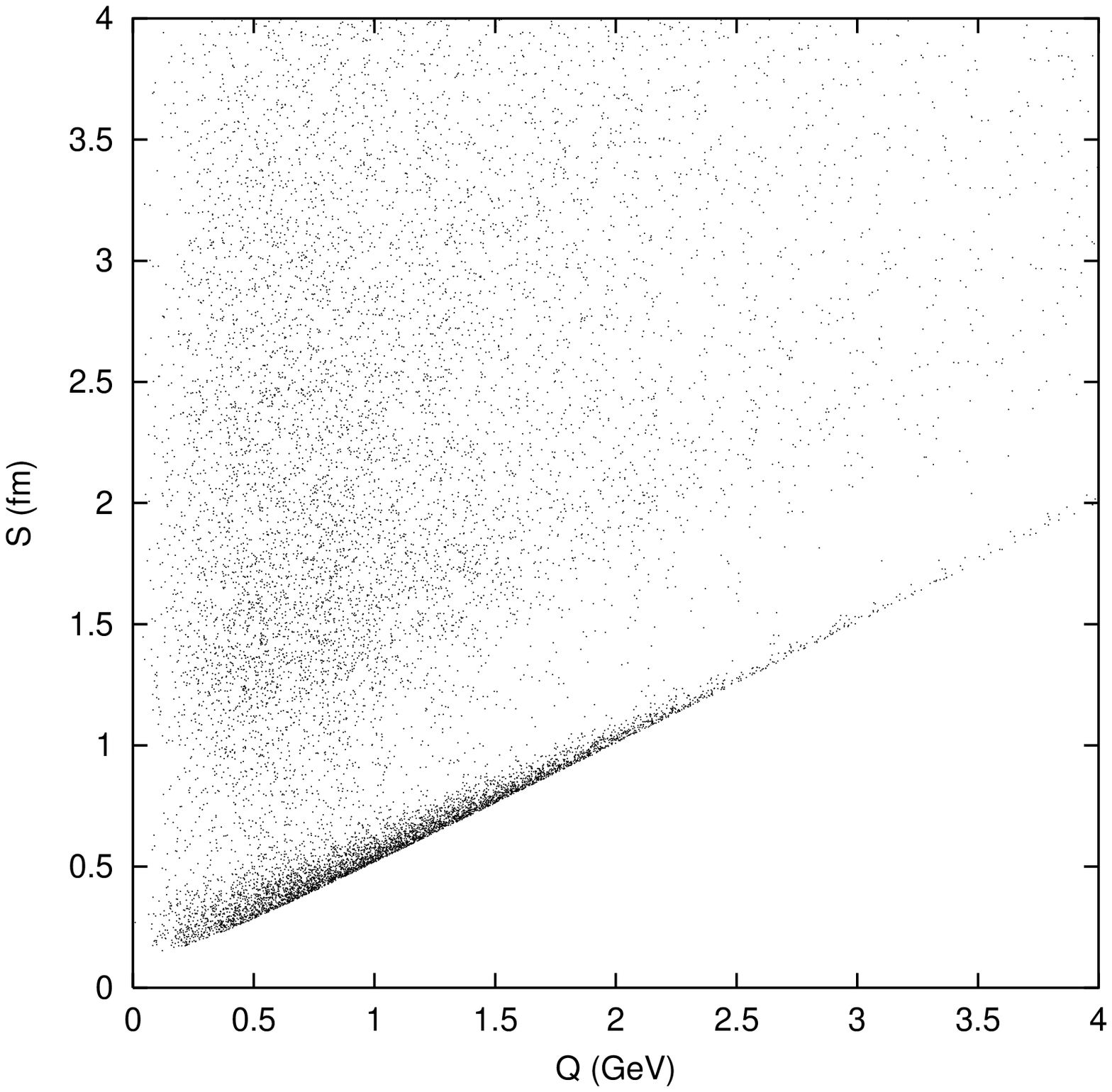, height=80mm}
\caption{The phase-space of $\pi^0$-pairs in 2-parton events ($\mathrm{u\bar{u}}$) at $E_\mathrm{cm}=91$ GeV. Right frame is a blow-up of the interesting region.} 
\label{fig:pi0} 
\end{figure}

\begin{figure}
\hspace*{-0.5cm}
\epsfig{file=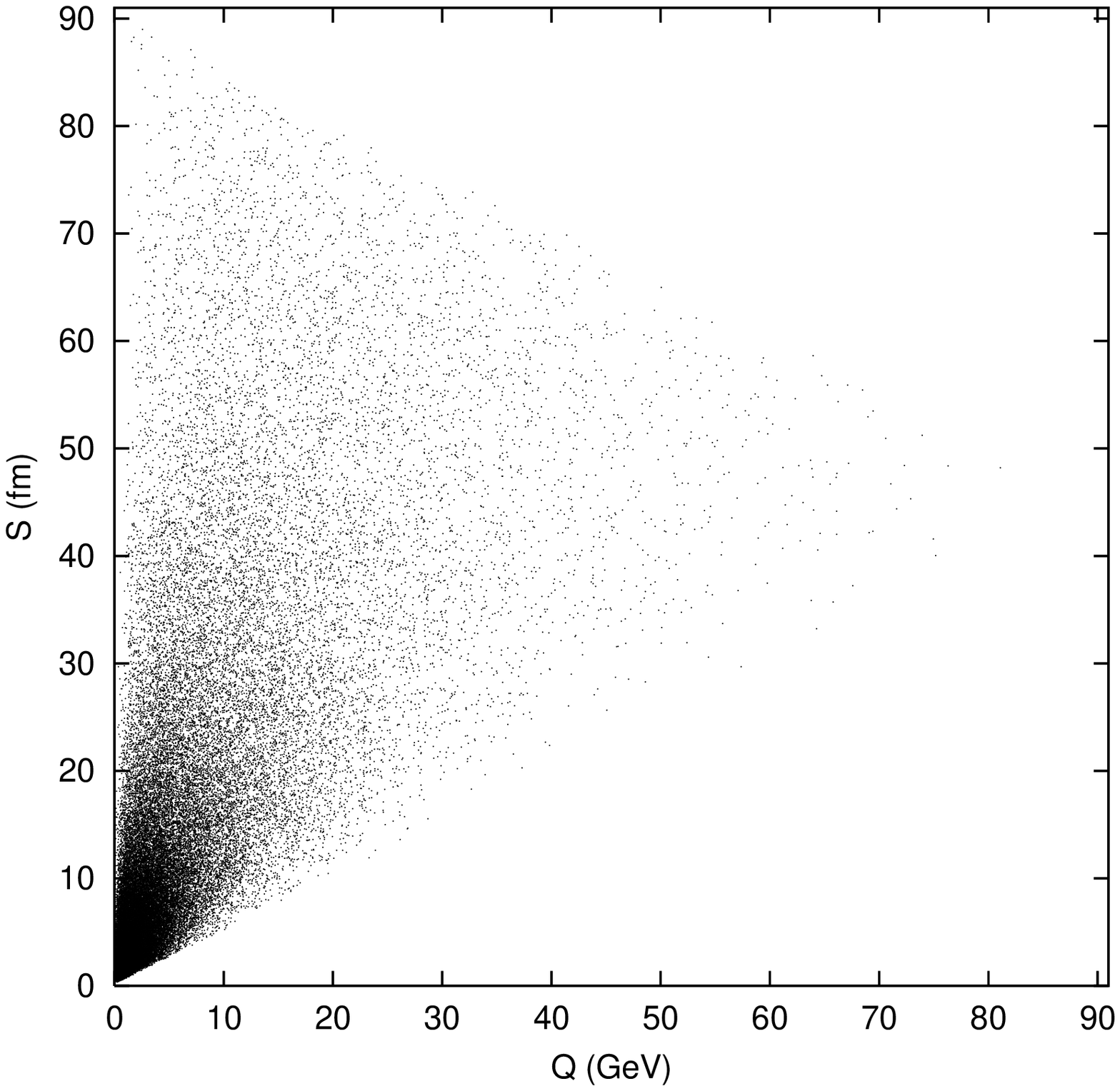, height=80mm}
\hspace*{-0.5cm}
\epsfig{file=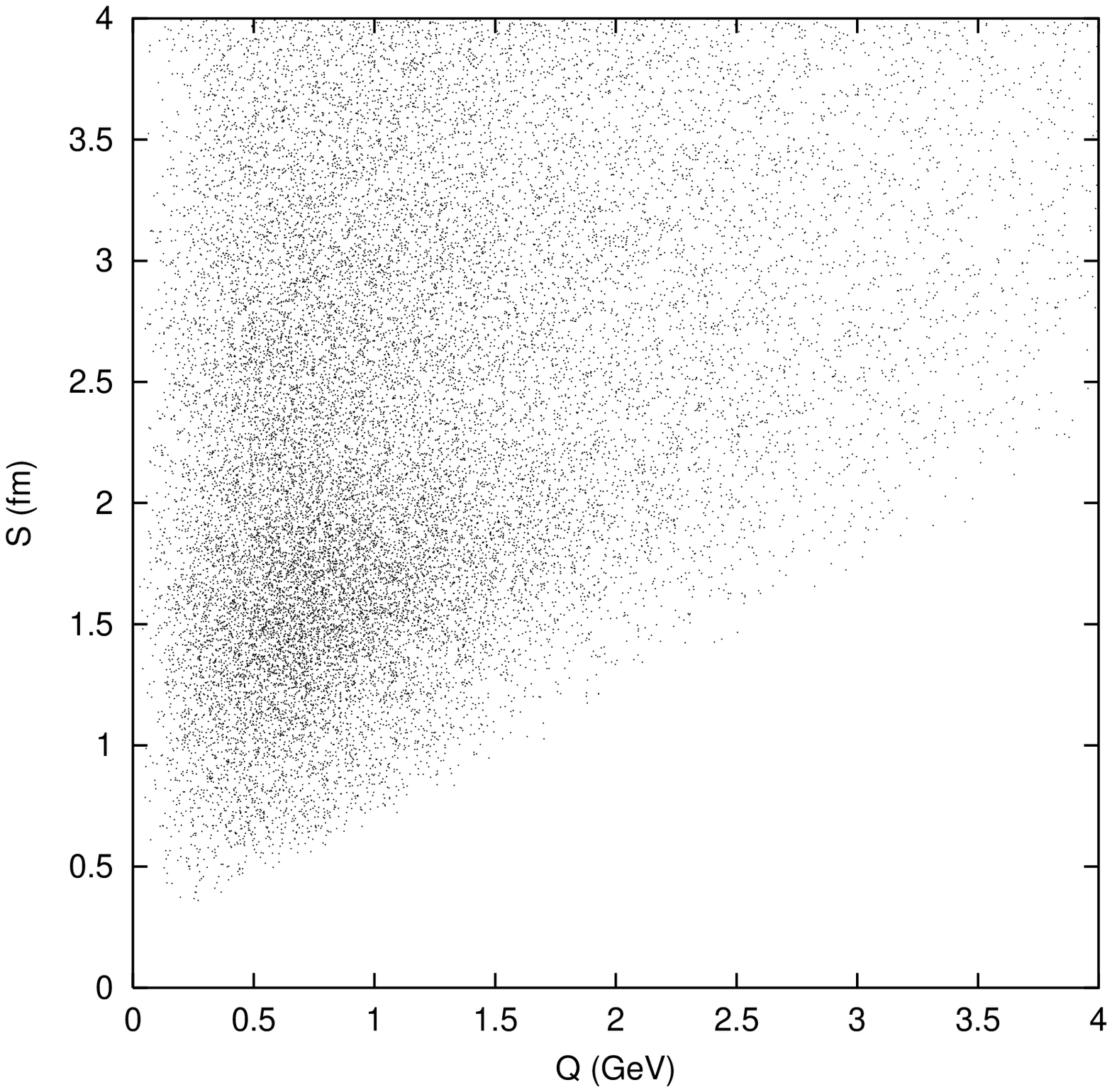, height=80mm}
\caption{As in Fig.~\ref{fig:pi0} but for $\pi^+$-pairs and $\pi^-$-pairs. The number of events plotted is the same as used in Fig.~\ref{fig:pi0}} 
\label{fig:pipm} 
\end{figure}

In order to invoke the Bose--Einstein correlation given by eq.~(\ref{cspatial}) we need to know how the phase-space looks like in the parameters $Q$ and $S$. It is interesting to see the phase-space of the basic event, the 2-parton event, even though we will not make further use of it. This is shown by the \textsc{Pythia}-generated scatter plots in Fig.~\ref{fig:pi0} and Fig.~\ref{fig:pipm} for the two most common mesons. At first the $Q$--$S$ dependence is not as clear as one might expect, but it is obvious that the phase space is restricted into a triangular area, of a size that depends on the system rest energy $E_\mathrm{cm}$, and a shape that depends on the definition of the hadron production vertex. It is mainly the phase-space restrictions and the fact that low $Q$ and $S$ seems favoured that give rise to the behaviour of the mean $\langle S \rangle$, as a function of $Q$, shown in Fig.~\ref{fig:plot2}. The phase--spaces of $\pi^0$-pairs and $\pi^\pm$-pairs are similar at large scales, but there is one difference that might become very important for their Bose--Einstein behaviour. Neutral pions pairs are closely clustered along the lower bound in $S$, which corresponds to nearest neighbour pairs, whereas charged pions are forbidden to be produced in such pairs. The same effect shows up in the less interesting upper bound of $S$: the flavour-order in which hadrons are produced in the 2-parton case prohibits charged identical particles from being produced at the two string ends. 

For multiparton events the phase-space will not look as simple as above, instead it will more and more take on a familiar shape coming from randomly distributed pairs in real and momentum space. Thus we have to continue our discussion assuming a small correlation in $Q$ and $S$.
  
If we wish to follow the path taken in section~\ref{sec:luboei} in a more general manner we would look for a shift $(\delta_Q,\delta_S)$ in both variables. This would correspond to a situation where we were given a distribution in $(Q,S)$, in a world without Bose--Einstein effects, and then were asked how this distribution with a minimal shift\footnote{A minimal shift is a shift that alters the physical picture (string geometry etc.) the least, but still gives the correct distribution} $(Q,S) \rightarrow (Q,S) -(\delta_Q,\delta_S)$ would be turned into the correct Bose--Einstein distribution.  In the situation where the difference in momentum and position for identical pairs of bosons are isotropically and homogeneously distributed throughout both spaces the ratio between the correct and the non-Bose--Einstein distribution is given by the correlation function in eq.~(\ref{cspatial}), according to our assumption. 

A two-parameter shift would only work if $Q$ and $S$ are independent variables, however, which is not entirely true. A shift in $Q$ would induce an indirect shift in $S$ and vice versa, hence this two-parameter shift is problematic. We have therefore chosen only to alter the $Q$-distribution, at the expense of not performing a minimal shift, i.e. one might choose a larger $\delta_Q$ instead of smaller $(\delta_Q,\delta_S)$. Of course there is still an indirect shift in $S$, that we will ignore. That is, given an original event without Bose--Einstein effects, the $S$, $Q$ and $\delta_Q$ are evaluated, but the procedure is not iterated to provide updated $S$ values. Furthermore, $Q$ is a measurable quantity, which is not true for $S$. 

We do not know for sure how the shift of momenta from Bose--Einstein effects are correlated with shifts in the space--time picture, so, in order to proceed, we make the crude assumption that $S$ is constant under a shift in $Q$, thus giving a constant contribution to the phase-space density, which factors out. Then nothing changes from the derivation of $\delta (Q)$ in section~\ref{sec:luboei}, except for the parameterization of the correlation function, and one obtains a shift in the $Q$-variable given by:
\begin{equation}
\delta (Q,S)=\lambda \frac{\sqrt{4m^2 +Q^2 }}{Q^2} \int^Q_0 
\frac{q^2}{\sqrt{4m^2 +q^2 }} \mathrm{e}^{-(kqS)^{\eta}}\mathrm{d}q
\end{equation}
The integral is not easily solved in an analytically useful form. Instead we had better perform a numerical integration in a computer efficient way whenever implementing it.   

\section{Results}

We have evaluated the algorithms of the proposed new Bose--Einstein model using the event generator \textsc{Pythia}. Since the particular choice of phenomenological Bose--Einstein model ultimately depends on experimental comparisons our objective is to simulate a realistic experimental situation and to produce data that are as close as possible to the one obtained in the real world. But since relevant experimental data are scarce, and since comparisons are best performed by experimentalists, we will only qualitatively compare the new $(Q,S)$-dependent algorithm with the current Lund model $Q$-dependent Bose--Einstein algorithm. To simplify the discussion we choose to denote the two models with $\mathrm{BE_{QS}}$ and $\mathrm{BE_{Q}}$, respectively, and NoBE is used to denote a reference null-model, which is the Lund Model without Bose--Einstein effects. 

The $\mathrm{BE_{Q}}$-version used is the one with global energy compensation, which is referred to as $\mathrm{BE_{0}}$ in~\cite{pythia}. In the $\mathrm{BE_{QS}}$  algorithm we use the same kind of energy compensation. Whenever comparing $\mathrm{BE_{Q}}$ and $\mathrm{BE_{QS}}$ we will always set the switches equivalently, so that the models only differ by the $S$-parameterization in the correlation function, e.g. when comparing exponential models we set $\eta=1$ and $\mathrm{MSTJ(51)}=1$, when comparing Gaussian models we set $\eta=2$ and $\mathrm{MSTJ(51)}=2$. Note: When developing the $\mathrm{BE_{QS}}$ model we had the idea that $S \propto Q$ for small $Q$'s, thus one might want to compare $\mathrm{BE_{QS}}$(exp) with $\mathrm{BE_{Q}}$(Gauss), since the the two correlation functions show similar $Q$-dependence. But, in fact, $S$ approaches asymptotically a constant ($\sim$ 1 GeV) in the low-$Q$ limit, and the above stated comparisons are more relevant. 

We will only discuss the lightest and most abundant meson, the pion, of which like-charged pion-pairs ($\mathrm{\pi^+\pi^++\pi^-\pi^-}$) have the same Bose--Einstein behaviour, which is slightly different from the one shown by neutral pion-pairs ($\mathrm{\pi^0\pi^0}$). Since charged pions are more often used in experiments, being easier to detect and measure due to their charge, we will focus on the behaviour of these particles, and the data from neutral pions will be presented when the results are interesting enough. 

In the analysis we used the subroutine PYEEVT that generates random $n'$-parton events according to realistic $\mathrm{e^+e^-}$ annihilations at LEP1, at $E_\mathrm{cm}=91$ GeV. These were classified to $n$-jets using the subroutine PYCLUS (Scaled Durham distance: MSTU(46) $=$ 6, Maximum joining distance: $y=$ PARU(45) $=$ 0.005~\cite{pythia}) such that the variable number $n$ was obtained. For statistical reasons all events with $n\geq 6$ were added to the 5-jet events. 

Here it becomes obvious why we choose to make the distinction between an $n'$-parton event and an $n$-jet. The number of partons $n'$ in an event is not an experimentally measurable quantity; furthermore, in principle, but not manifestly included in PYEEVT, there are an infinite number of partons in the limit where the gluon energy or emission angle goes to zero. For practical purposes all one can do is to define a suitable experimental criterion, as done above, that allows one to deduce the number of distinguishable jets $n$ out of a larger number of indistinguishable partons $n'$.
\begin{figure}
\begin{center}
\epsfig{file=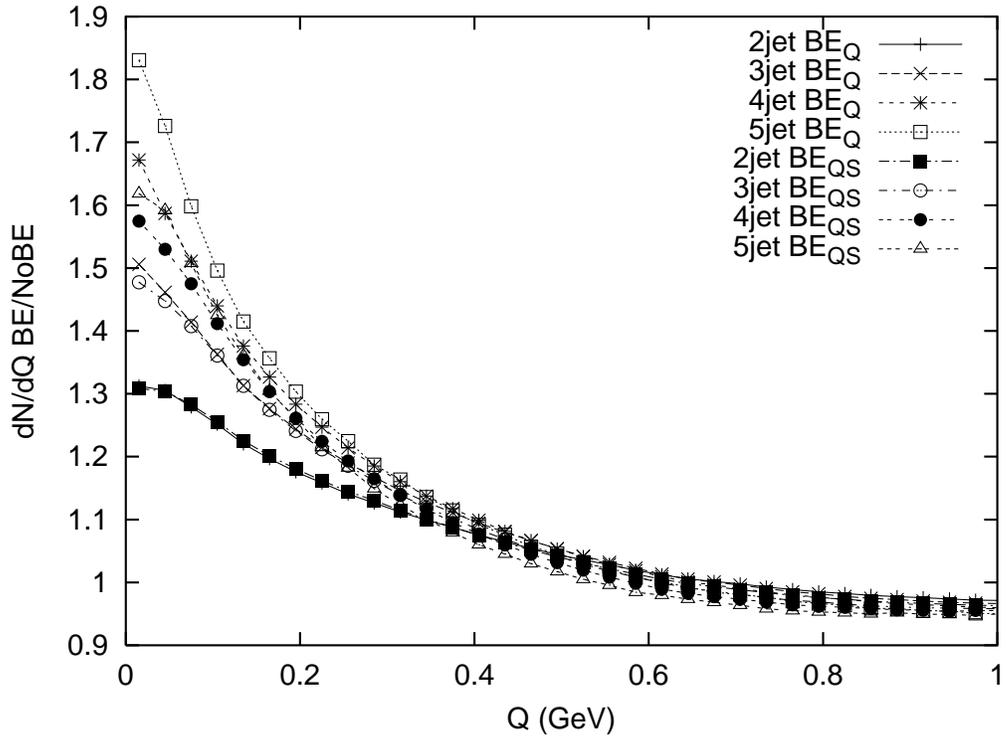, height=100mm}
\caption{The $\mathrm{BE_{QS}/NoBE}$ and $\mathrm{BE_{Q}/NoBE}$ ratios in the $Q$-distribution for charged pion-pairs. Exponential models are used.}  
\label{fig:be1} 
\end{center}
\end{figure}
\begin{figure}
\begin{center}
\epsfig{file=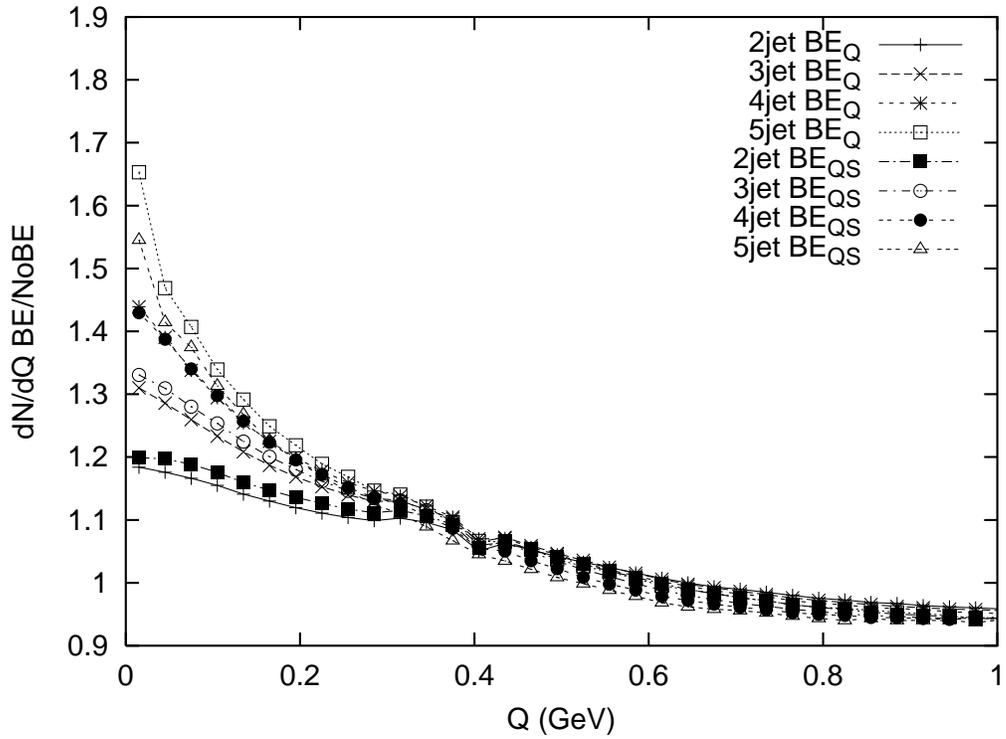, height=100mm}
\caption{As in Fig.~\ref{fig:be1} but for neutral pion-pairs. Exponential models are used.}  
\label{fig:be2} 
\end{center}
\end{figure}

For these simulated LEP1-events the $Q$-distribution $\mathrm{d}N/\mathrm{d}Q$, $N$ being number of pairs in an event, was collected for charged pion-pairs and neutral pion-pairs. This also includes pairs where one or both come from decays of long-lived particles and thus cannot display Bose--Einstein effects. The familiar Bose--Einstein ratio BE/NoBE, in the $Q$-distribution, was then obtained by: 
\begin{equation}
\frac{\mathrm{d}N_\mathrm{BE}}{\mathrm{d}N_\mathrm{NoBE}}=\frac{\mathrm{d}N_\mathrm{BE}/\mathrm{d}Q}{\mathrm{d}N_\mathrm{NoBE}/\mathrm{d}Q}
\end{equation}
This ratio, which should resemble the correlation function $c(Q)$, is shown in Fig.~\ref{fig:be1} and Fig.~\ref{fig:be2} for the two exponential models and for different jet multiplicities. Note that the number of $n$-jet events differ between the BE and NoBE alternatives, and that therefore the $\mathrm{d}N/\mathrm{d}Q$ distributions are normalized per event before taking ratios. 

\begin{figure}
\begin{center}
\epsfig{file=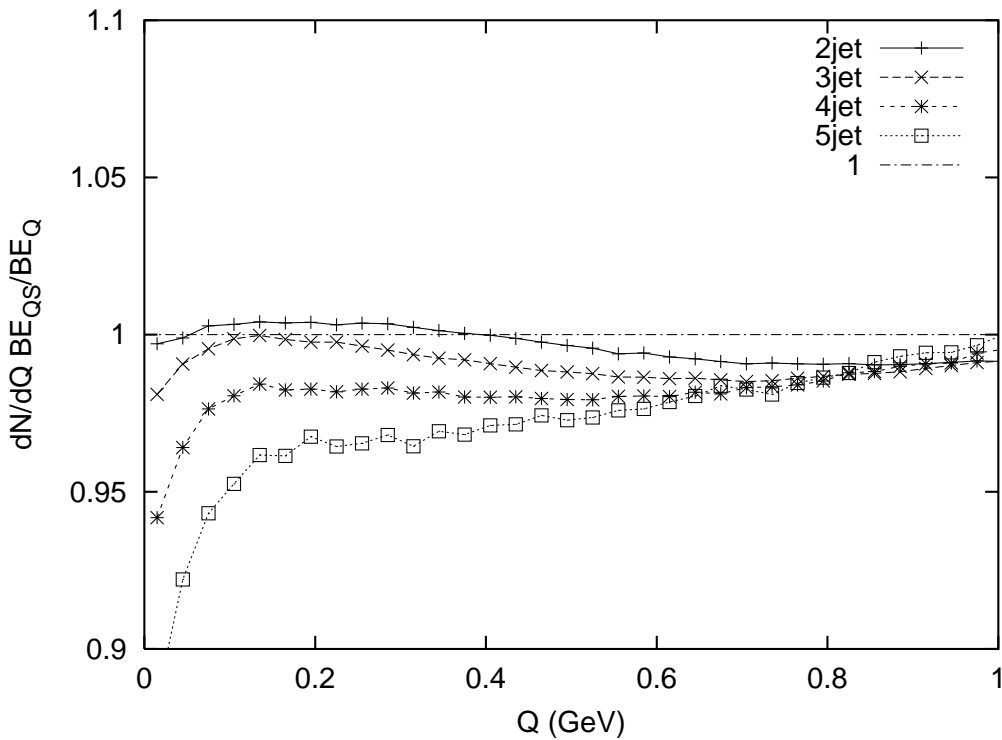, height=100mm}
\caption{The $\mathrm{BE_{QS}/BE_{Q}}$ ratios in the $Q$-distribution for charged pion-pairs. As guidance for the eye the constant 1 is plotted as well. Exponential models are used.}  
\label{fig:final} 
\end{center}
\end{figure}
\begin{figure}
\begin{center}
\epsfig{file=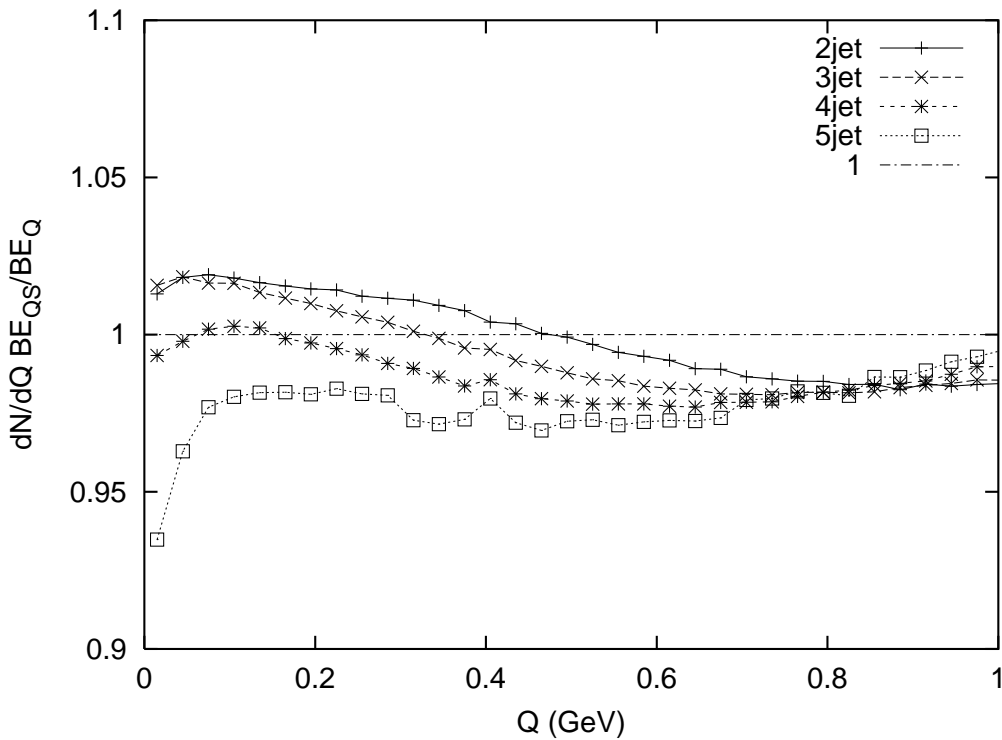, height=100mm}
\caption{As in Fig.~\ref{fig:final} but for neutral pion-pairs. Exponential models are used. }  
\label{fig:final2} 
\end{center}
\end{figure}

Considering charged pions and exponential models, Fig.~\ref{fig:be1} shows that the Bose--Einstein effect is stronger for higher jet multiplicities in both models, but that the new $\mathrm{BE_{QS}}$ model suppresses the effects relatively to the $\mathrm{BE_Q}$ model. Since the $\mathrm{BE_Q}$ model, as far as known, reproduce experimental data in the 2-jet case we have normalized the $\mathrm{BE_{QS}}$ model to give the same behaviour in this case for charged pions, by setting the free parameter $k$ in the correlation function~(\ref{cspatial}) to $k=1/11$ (for $\kappa=1$ GeV/fm and recalling that $\hbar=0.197327$~$\mathrm{GeV}\cdot \mathrm{fm}$). The value of $k$ is relatively small, one would perhaps expect a value close to unity, but it might reflect the fact that in our model we have only used the average position of a hadron at time of creation, not the hadron wavefunction. The wavefunction is of course spread out in space and the effects of a symmetrization is most notable at points in-between particles, thus corresponding to a smaller $S$ than we have used. The value of $k$ is also dependent on other parameters, in particular $R$ since we have fitted the $\mathrm{BE_{QS}}$ model to the $\mathrm{BE_{Q}}$ model, which depends on $R$. The relation is that $R$ and $k$ should be approximately proportional. The relevant parameters used throughout the analysis are: the constant $R$ in $\mathrm{BE_{Q}}$ correlation function~(\ref{first}) is set to $R=0.6$ fm; $\lambda$ in both correlation functions~(\ref{first},\ref{cspatial}) is set to $\lambda=1.35$. 

For the neutral pions the Bose--Einstein correlation is even more suppressed than for charged pions in both models. This somewhat unexpected behaviour has a simple explanation in the decays of resonances that produce a multitude of uncorrelated neutral pions that will dilute the real effect in our way of presenting data.   

The contrast is small between the two models in Fig.~\ref{fig:be1} and Fig.~\ref{fig:be2}, thus, instead we make use of a double ratio in order to enhance any differences:
\begin{equation}
\frac{\mathrm{d}N_\mathrm{BE_{QS}}}{\mathrm{d}N_\mathrm{BE_Q}}=\frac{\mathrm{d}N_\mathrm{BE_{QS}}/ \mathrm{d}N_\mathrm{NoBE}}{\mathrm{d}N_\mathrm{BE_Q}/ \mathrm{d}N_\mathrm{NoBE}}
\end{equation}
This is shown in Fig.~\ref{fig:final}, where one for charged pions can see a relative suppression in the Bose--Einstein correlation of the $\mathrm{BE_{QS}}$ model. For 4-jets there is a 2--3 percent suppression and for 5-jets and higher there is a 4--5 percent suppression depending on the choice of $Q$-region. In the case of neutral pions, Fig.~\ref{fig:final2} show a similar behaviour between different jets, but there is also an overall shift upwards. Note that the general level of Bose--Einstein effects always can be adjusted by a parameter retuning, so it is the trend with increasing number of jets that is the interesting feature.
\begin{figure}
\begin{center}
\epsfig{file=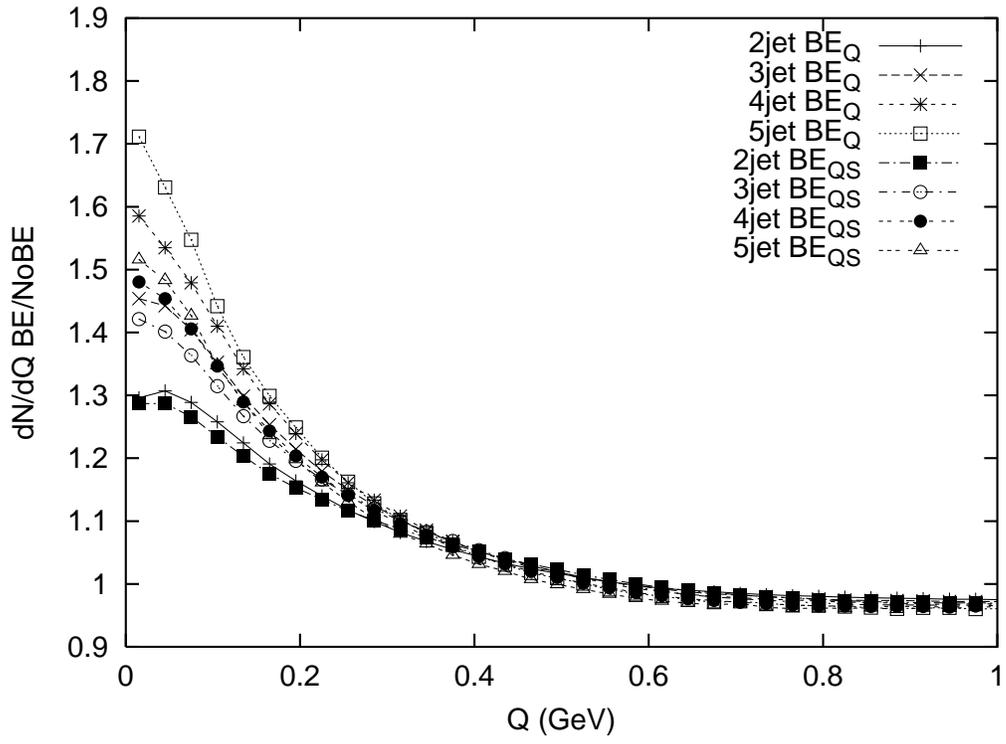, height=100mm}
\caption{As in Fig.~\ref{fig:be1} but for Gaussian models and charged pions.}  
\label{fig:be5} 
\end{center}
\end{figure}
\begin{figure}
\begin{center}
\epsfig{file=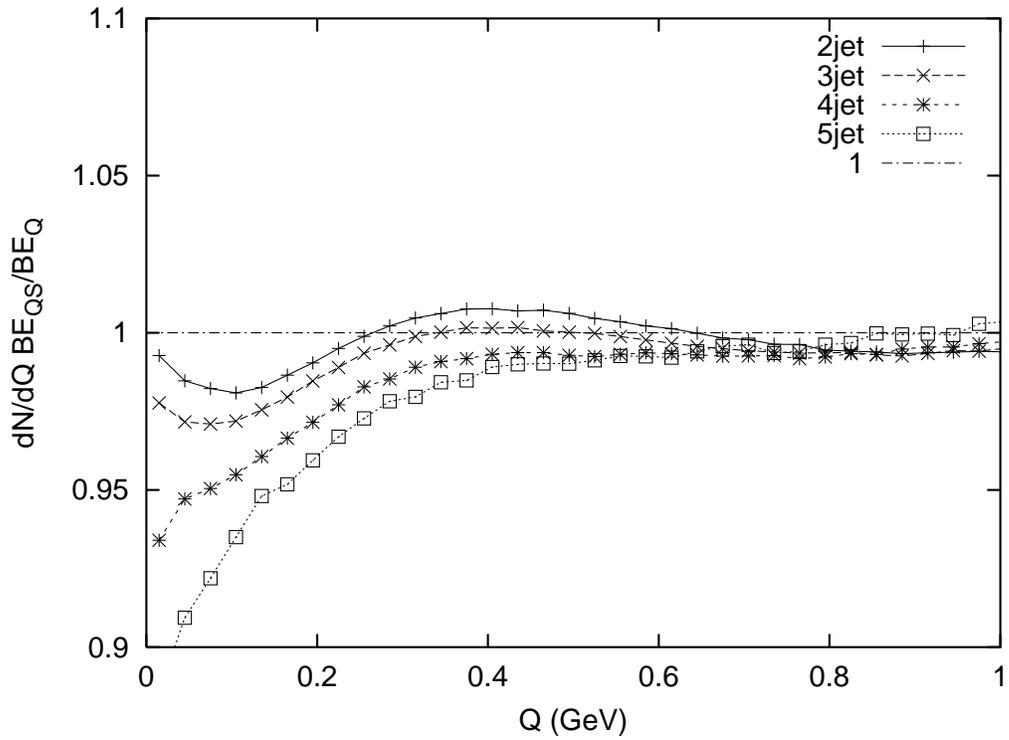, height=100mm}
\caption{As in Fig.~\ref{fig:final} but for Gaussian models and charged pions.}  
 \label{fig:final5} 
\end{center}
\end{figure} 

We have so far discussed the comparison between the two models with exponential correlation functions, but, in fact, quantitatively the results obtained with Gaussian correlation functions are very similar. In Fig.~\ref{fig:be5} the BE/NoBE ratios and in Fig.~\ref{fig:final5} the $\mathrm{BE_{QS}/BE_Q}$ ratios are given for charged pions. The shape of the curves in this last plot is somewhat different from the equivalent Fig.~\ref{fig:final} for exponential models. But the 2-jet fit is consistent with an approximative value of $k=1/9$ (used in all Gaussian plots), which is slightly larger than the one obtained from the exponential fit. Thus it seems that the Gaussian parameterization in itself gives stronger Bose--Einstein correlation. Which of the two different types of shapes that is more true is ultimately an experimental question.

For neutral pions with Gaussian models the result is, as expected, equivalent to earlier result: the Be/NoBE ratios are very similar to Fig.~\ref{fig:be2} (neutral pions, exponential), and the $\mathrm{BE_{QS}/BE_Q}$ ratios are very similar to Fig.~\ref{fig:final5} (charged pions, Gaussian) with an overall increase (above unity), which corresponds to stronger Bose--Einstein correlation with the new model. This we have already observed for exponential models in the neutral and charged pions plots.

Since it is sometimes possible in experiment to distinguish $\mathrm{e^+e^-} \rightarrow \mathrm{Z^0} \rightarrow \mathrm{u\bar{u}} / \mathrm{d\bar{d}}/\mathrm{s\bar{s}}$ events from heavy quark events, $\mathrm{e^+e^-} \rightarrow \mathrm{Z^0} \rightarrow \mathrm{c\bar{c}}/\mathrm{b\bar{b}}$, we have also looked at the former events specifically. No extra plots will be presented since the result does not differ substantially, but one can generally say that the Bose--Einstein effect in this case is less diluted. There is approximately a 15 percent increase in all BE/NoBE ratios, but for the $\mathrm{BE_{QS}/BE_Q}$ ratios the change is not significant. 

It should be noted that for all plots with the ratio $\mathrm{BE_{QS}/BE_Q}$, i.e. Fig.~\ref{fig:final}, Fig.~\ref{fig:final2} and Fig.~\ref{fig:final5}, the ratios do not approach unity outside the range included in plots ($Q>1$). Most of the ratios take some constant value in the interval [1.00,1.02] for $Q>2$. However, it is not required that this ratio should become unity in an event generator, since for both models $\mathrm{d}N/\mathrm{d}Q$ approach zero very fast in this range, and small effects are blown up.

\subsection{The effect of the $\mathrm{BE_{QS}}$ algorithm in different energy bins}

\begin{figure}
\begin{center}
\epsfig{file=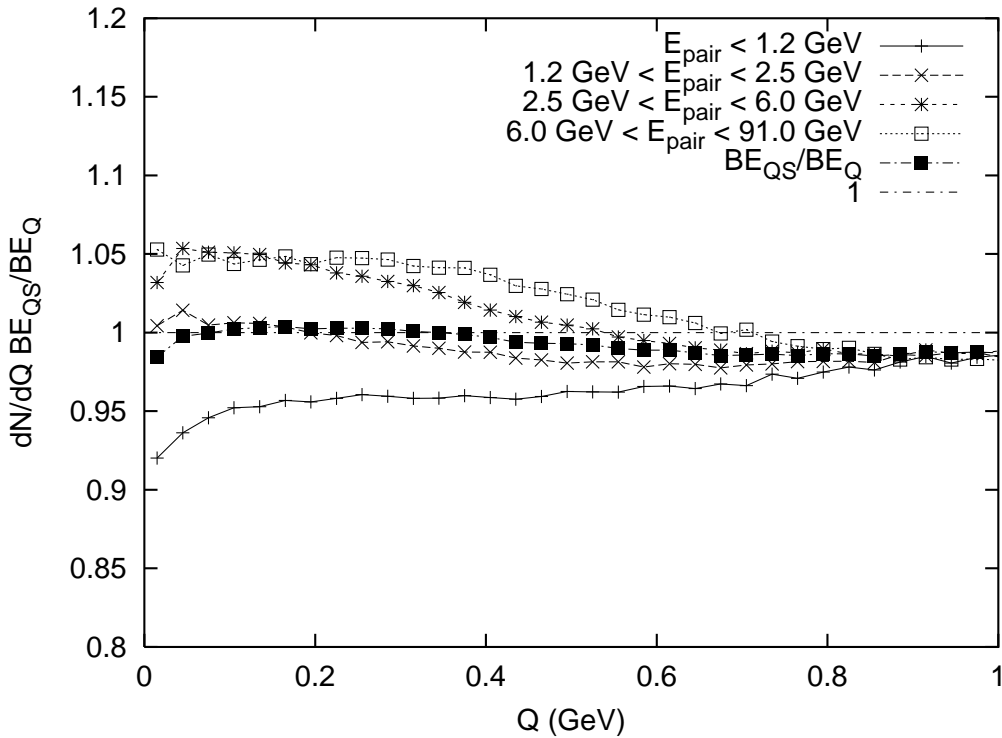, height=100mm}
\caption{The $\mathrm{BE_{QS}/BE_{Q}}$ ratios in the $Q$-distribution for charged pion-pairs in the 2-jet case ($\mathrm{u\bar{u}}$). The intervals give the relevant $E_\mathrm{pair}$-bins, whereas the curve labelled by $\mathrm{BE_{QS}/BE_{Q}}$ gives the total ratio for 2-jets, which also occurs in Fig.~\ref{fig:final}, but there for all quark flavours. Exponential models are used.}
\label{fig:dubb2jet} 
\end{center}
\end{figure}
\begin{figure}
\begin{center}
\epsfig{file=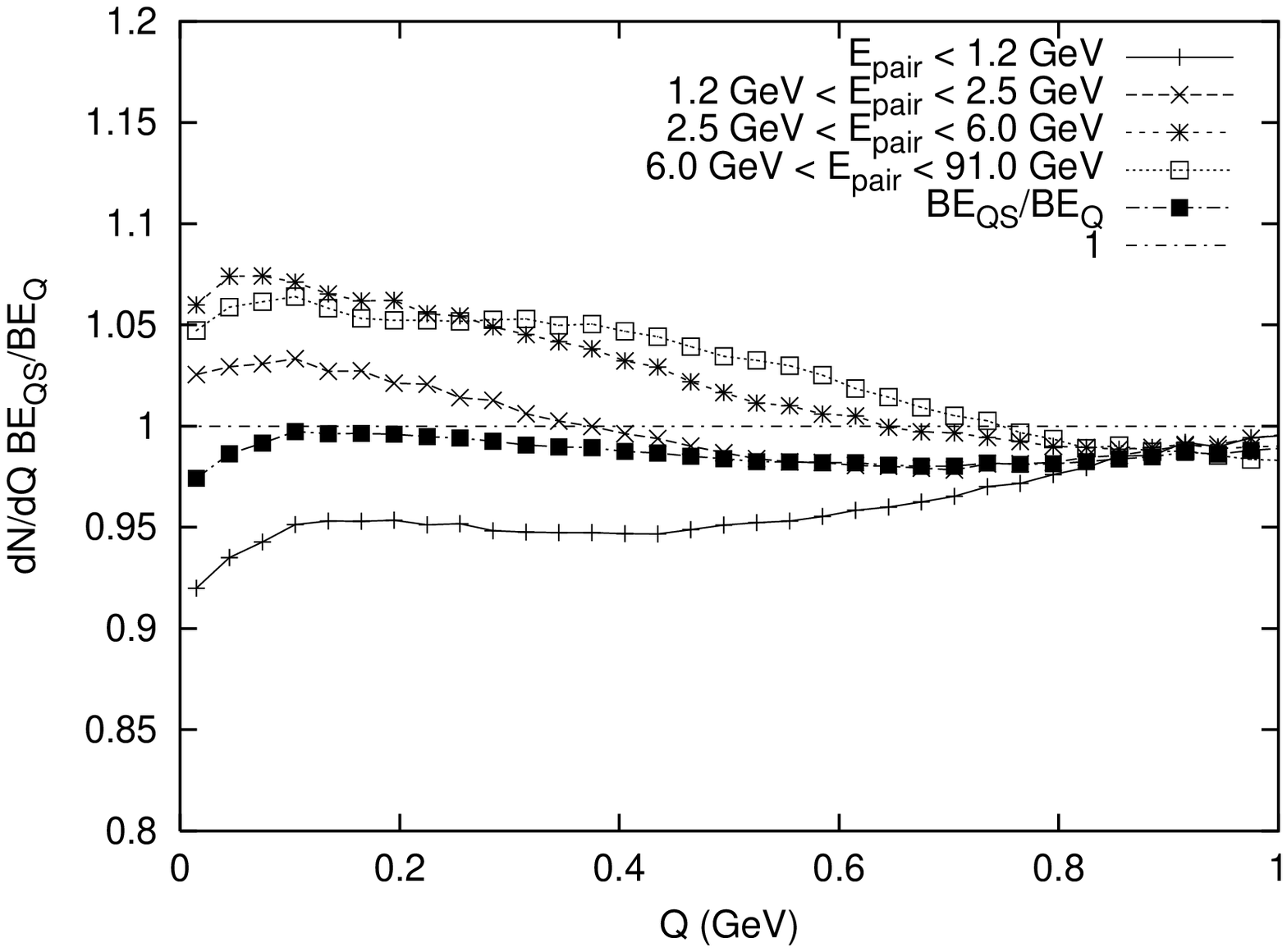, height=100mm}
\caption{As in Fig.~\ref{fig:dubb2jet} but for 3-jets.}
\label{fig:dubb3jet} 
\end{center}
\end{figure}
\begin{figure}
\begin{center}
\epsfig{file=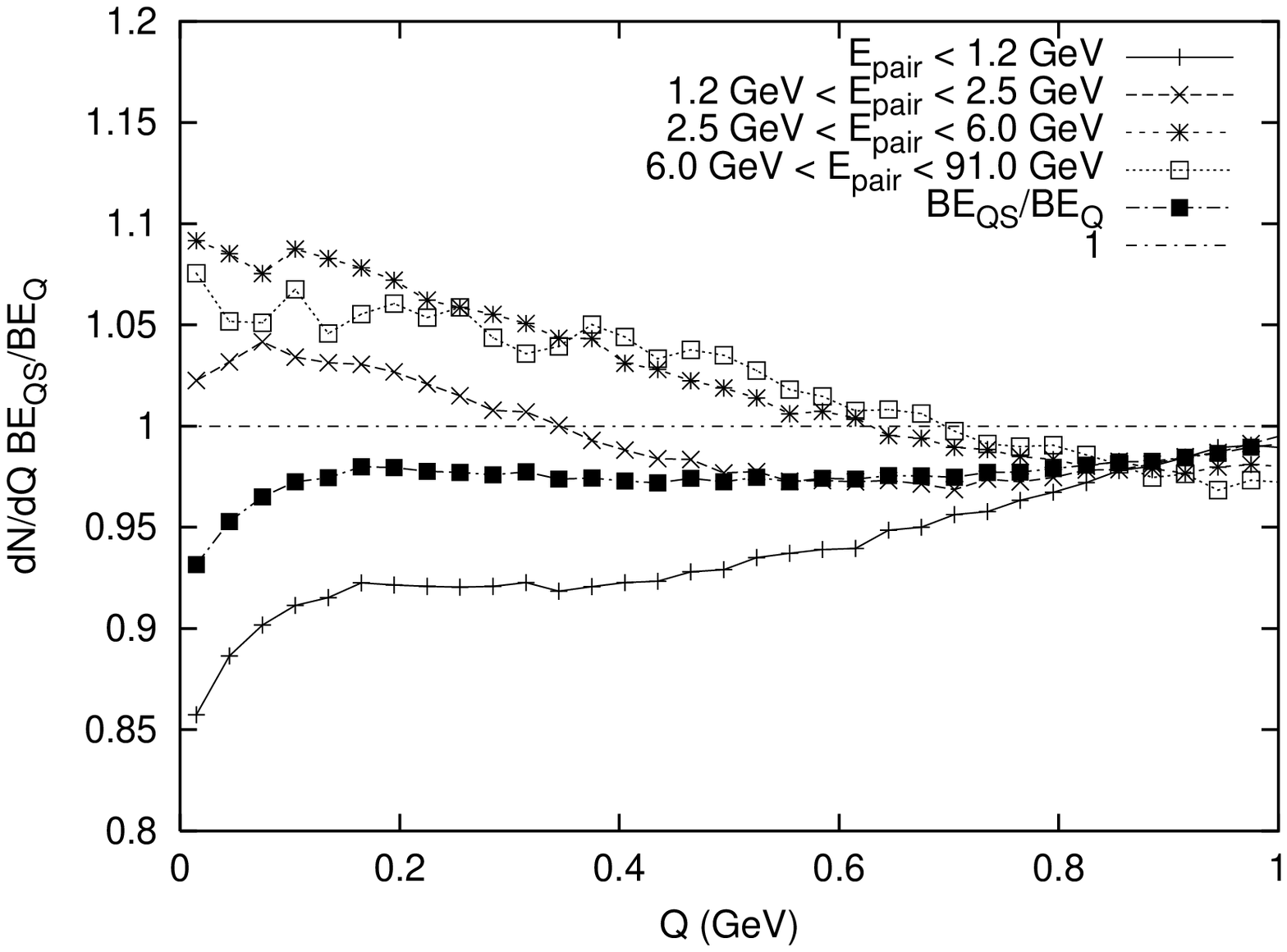, height=100mm}
\caption{As in Fig.~\ref{fig:dubb2jet} but for 4-jets.}
\label{fig:dubb4jet} 
\end{center}
\end{figure}
\begin{figure}
\begin{center}
\epsfig{file=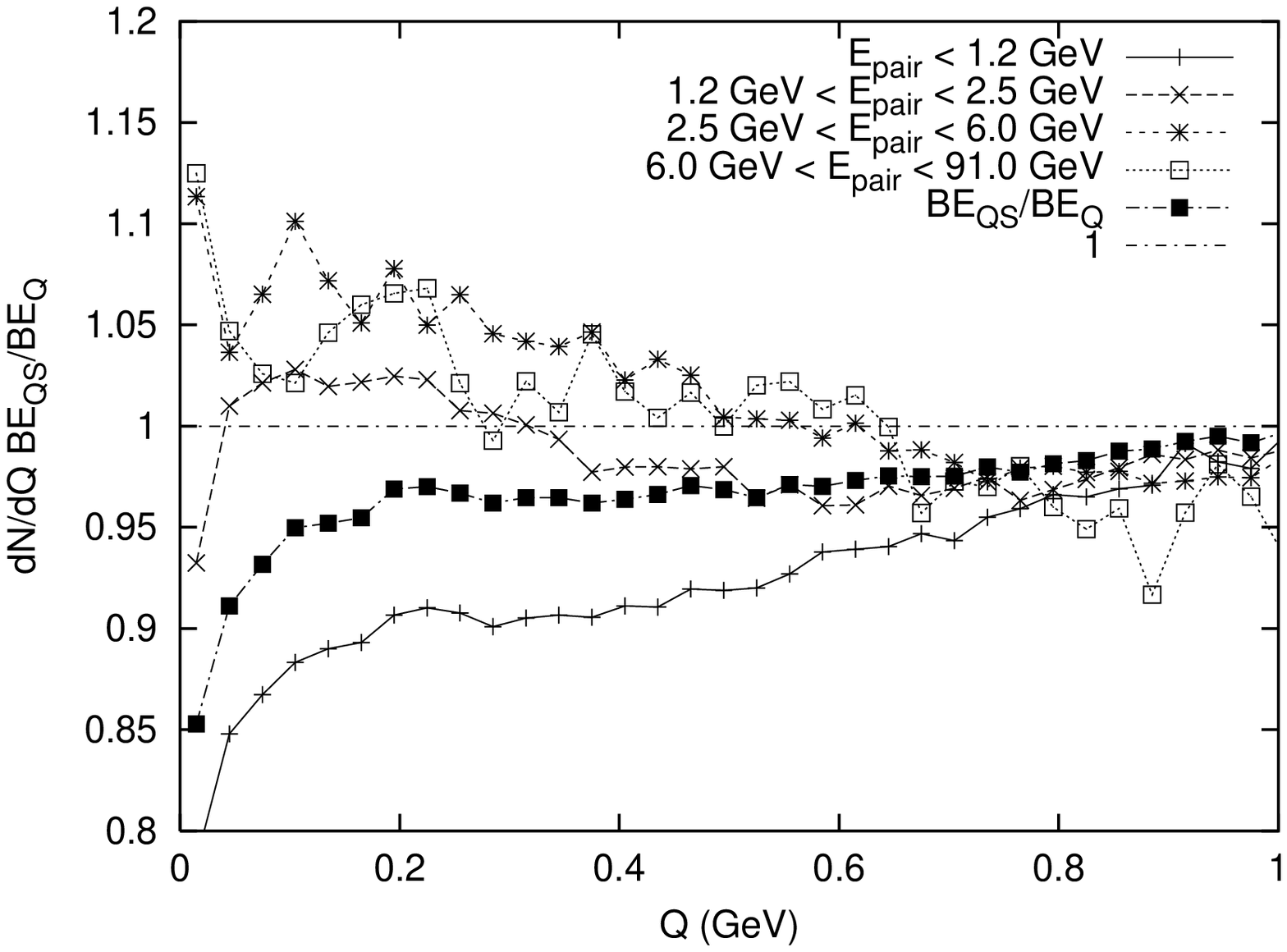, height=100mm}
\caption{As in Fig.~\ref{fig:dubb2jet} but for 5-jets and higher orders. Note: Fluctuations are statistical, not caused by model.}
\label{fig:dubb5jet} 
\end{center}
\end{figure}

The smallness of the BE-suppression for charged pions obtained with $\mathrm{BE_{QS}}$ is impressive but discomforting: impressive that so stable, i.e. the naive $\mathrm{BE_Q}$ is not that bad, but discomforting since it becomes hard to tell if the suppression is significant enough. This leads us into further investigations of the actions of this new algorithm. A critical test is whether or not the new algorithm fulfils the following intuitive insight. For higher multiplicities, say 4-jets, the string regions are relatively well separated in space, but the central parts of each region holds relatively little of the parton momentum, thus these parts are closer in momentum space. Hadrons coming from these central parts have thus low $Q$ and high $S$, which should enhance differences in a comparison between $\mathrm{BE_Q}$ and $\mathrm{BE_Q}$, if the new algorithm satisfy our expectations. More specifically, we want to compare meson-pairs that inherit a small portion of the parton momenta, i.e. mesons in the centre of a 2-parton region, to pairs that inherit a large portion of the parton momenta, i.e. mesons closer to the partons in the space--time picture.   

In order to distinguish such pairs we choose criteria that are also useful for experimentalists. For two mesons with energy $E_1$ and $E_2$ in the total system rest frame, we define $E_\mathrm{pair}=E_1+E_2$. Then we create four energy-bins by cutting the allowed $E_\mathrm{pair}$ range into appropriate regions, given in Fig.~\ref{fig:dubb2jet} to Fig.~\ref{fig:dubb5jet}. These plots show the double ratio $\mathrm{BE_{QS}/BE_{Q}}$ for each separate $E_\mathrm{pair}$-bin. 

By splitting the data into four regions like this we will introduce some unavoidable correlations, since $Q$ and $E_\mathrm{pair}$ are not independent quantities. But the effect will be small unless $Q$ is near a cut in the $E_\mathrm{pair}$-range. It should be noted that the relative number of pairs in an $E_\mathrm{pair}$-bin vary from 2-jet to 5-jet, thus influencing the relative behaviour of the plots when comparing different jets in Fig.~\ref{fig:dubb2jet} to Fig.~\ref{fig:dubb5jet}.

If pairs with low-$E_\mathrm{pair}$ are interpreted as hadrons that have been created in the central parts of parton-regions, then it is clear that $\mathrm{BE_{QS}}$ model will strongly suppress Bose--Einstein correlations from these pairs, e.g. 5 percent for 2-jets and 10 percent for 5-jets and higher. But, oppositely, high-energetic pairs, corresponding to hadrons created near region boundaries, will with the same model contribute with an intensified Bose--Einstein effect of similar strength. However, the net effect will always be a suppression for $n$-jets with $n>2$, as already noted. Since the $\mathrm{BE_{QS}}$ algorithm is roughly tuned to have the same overall $\mathrm{d}N/\mathrm{d}Q$ distribution as $\mathrm{BE_{Q}}$, the supression of Bose--Einstein effects at small $E_\mathrm{pair}$ in the former then leads to an enhancement at large $E_\mathrm{pair}$. Had the tune been for agreement at large $E_\mathrm{pair}$, the drop at small $E_\mathrm{pair}$ would have been even more dramatic, up to 20 percent for  4-jet events. 

We present only plots for charged pions and exponential models, with initial u-quarks, i.e. only $\mathrm{e^+e^-} \rightarrow \mathrm{Z^0} \rightarrow \mathrm{u\bar{u}}$ events (this increases $E_\mathrm{pair}$-bin spread by approximately 60 percent compared to the full mixture of quark flavours). Neutral pions-pairs behave essentially the same way as charged do in Fig.~\ref{fig:dubb2jet} to  Fig.~\ref{fig:dubb5jet}. Of course, there is a general shift upwards in the curves corresponding to these particles, which can be seen already in Fig.~\ref{fig:final2} versus Fig.~\ref{fig:final}. In addition, for neutral pions there is a smaller spread between the curves coming from different $E_\mathrm{pair}$-bins. The spread is about 85 percent of the spread seen for charged pions. However, this last effect is probably caused by the dilution effect from decays, mentioned earlier, that is stronger for neutral pions compared to charged pions.

Also when comparing the Gaussian models the result is the same as obtained with exponential models. It is always the low-$E_\mathrm{pair}$ bin that drops down by a few percent compared to collective curve, and high-$E_\mathrm{pair}$ rise by approximately the same percentage. The spread in $E_\mathrm{pair}$-bins is approximately 75 percent and 60 percent for charged and neutral pions, respectively, compared to the spread seen for charged pion with exponential models.

\section{Summary \& Outlook}

There has been an indication that the LUBOEI algorithm overestimates the Bose--Einstein effects for 4-jet events, when the parameters $R$ and $\lambda$ is tuned for 2-jets~\cite{kjaer}. We have reason to believe that this phenomenon has to do with the fact that the space--time picture is neglected in the correlation function $c(Q)$. This led us to try out the most simple extension of this model, by incorporating the space--time picture in one variable $S$, corresponding to an invariant separation of production vertices, thereby giving a correlation function $c(Q,S)$. 

The Lund Model space--time picture used for string fragmentation provides natural hadron production regions, where a single hadron may be associated with a point that represents the hadron production vertex. 

Including decays of primarily produced resonances and multi-parton events showed that the quantity $S$ cannot simply be the ordinary 4-vector distance of two hadron production vertices, since there can always be vertices that are at time-like separations as well as space-like. Instead a possible choice is that $S$ is the 3-space distance between the two hadrons in the pair rest frame when the late particle is produced. 

The structure of the Lund Model string allows an algorithm to untangle all the hadron production vertices from information given by hadron rank, hadron 4-momenta and parton 4-momenta. Unfortunately ambiguities in the hadronization algorithm occasionally gives hadron 4-momenta that cannot consistently be used for assigning production vertices with this algorithm. An easy but surprisingly functional solution to this problem is to evaluate a hadron production vertex in all possible parton regions and then choose the vertex that maximizes its proper time. 

The effect of the new correlation function depends on the phase-space given in variables $Q$ and $S$, thus the result will depend on how we choose to perform local shifts in these variables. Due to uncertainties we made the crude assumption that the shift is independent of the $S$ phase-space density and that the direct shift is only to be performed in the observable $Q$. 

The new $\mathrm{BE_{QS}}$ algorithm then proved to possess many desirable properties. For charged pions the Bose--Einstein effect was suppressed in 4-jets and 5-jets compared to the $\mathrm{BE_Q}$ algorithm, whereas 2-jets and 3-jets gave comparable results. This is qualitatively what is found in the preliminary LEP1-results of~\cite{kjaer}. For neutral pions the result was also an enhancement of the Bose--Einstein effect in general. However, in order for them to produce the same result for 2-jets the parameter $R$ has to be tuned separately for charged and neutral pions, with the relation $R_{\pi^\pm}>R_{\pi^0}$. And indeed this is consistent with experimental results~\cite{amsterdam}. 

In the new model we introduced a free parameter $k$, with the rough value 1/11, that was fixed by fitting the $\mathrm{BE_{QS}}$ 2-jet curve for charged pions to the equivalent $\mathrm{BE_{Q}}$ curve. 

We also found that the new algorithm acts differently depending on the energy of the pion-pair, $E_\mathrm{pair}$. The Bose--Einstein effect is relatively suppressed in the low-$E_\mathrm{pair}$ region, which accounts for the majority of pairs, but enhanced in the high-$E_\mathrm{pair}$ regions compared to the results given by $\mathrm{BE_{Q}}$.

The results so far are very interesting and we suggest that further work could be done by considering:

\begin{itemize}
\item{Study the effects of the new algorithm in $\mathrm{e^+e^-} \rightarrow \mathrm{W^+W^-}$ events, which give two quark--antiquark systems. These events have lately been given much attention in experimental studies, and they are very important indicators that show if the $\mathrm{BE_{QS}}$ algorithm is on the right track. The issue is whether or not there is an inter-string Bose--Einstein effect. If there is not, then the two string pieces are independent systems (no cross-talk), if there is, then it cannot be modelled by a simple $Q$-dependent model, the spatial inter-string distance must show up somehow in a realistic parameterization.}
\item{Different parameterization of the correlation function, such as using an 4-product between a spatial quantity and a momentum quantity in the exponent in eq.~(\ref{cspatial}).} 
\item{Modelling an elongation of the source in the same correlation function as suggested by experimental data~\cite{amsterdam}. We have not studied it, but the $\mathrm{BE_{QS}}$ model might already have such properties.}
\item{Study the phase-space density dependence on $S$ and take local shifts in both $Q$ and $S$ into considerations when producing a Bose--Einstein influenced $(Q,S)$-distributions.}
\item{Study the role of different ways to include secondary decays, i.e. method 1 versus 2 in section~\ref{sec:prod} and other possibilities.} 
\end{itemize}

\section{Acknowledgments}
I would like to thank all people at the Department of Theoretical Physics in Lund, and especially my supervisor Torbj\"orn Sj\"ostrand who came up with the idea for this thesis and guided me in the Lund Model domain.


\begin{thebibliography}{99}

\bibitem{string}
B.~Andersson, G.~Gustafson, G.~Ingelman and T.~Sj\"ostrand,
Phys.\ Rep.\  {\bf 97} (1983) 31.

\bibitem{luboei}
L.~L\"onnblad and T.~Sj\"ostrand,
Phys. Lett. B {\bf 351} (1995) 293;
Eur.\ Phys.\ J. C  {\bf 2} (1998) 165.
\bibitem{hbt}
R.~Hanbury Brown and R.~Q.~Twiss,
Phil. Mag. {\bf 45} (1954) 663; 
Nature {\bf 178} (1956) 1046.

\bibitem{goldhaber}
G.~Goldhaber, S.~Goldhaber, W.~Lee and A.~Pais,
Phys.\ Rev.\  {\bf 1} (1960) 300.

\bibitem{kjaer}
N.~van~Remortel et al.,
DELPHI 2002-032 CONF 566 (2002)

\bibitem{kane}
G. Kane,
`Modern Elementary Particle Physics' (ISBN 0-201-62460-5, Perseus Publishing, 1993).

\bibitem{desy}
T.~Sj\"ostrand,
Nucl. Phys. B {\bf 248} (1984) 469

\bibitem{bo}
B.~Andersson and W.~Hofmann,
Phys. Lett. B {\bf 169} (1986) 364;

B.~Andersson and M.~Ringn$\acute{\mathrm{e}}$r,
Nucl. Phys. B {\bf513} (1998) 627.


\bibitem{lorstad}
B.~L\"orstad,
Int. J. Mod. Phys. A {\bf4} (1989) 2861

\bibitem{delphi}
DELPHI Collaboration,
Phys. Lett. B {\bf 511} (2001) 159.

\bibitem{amsterdam}
S.Todorova--Nova, `BE correlations at LEP \& HERA', presented at ICHEP 2002, Amsterdam (2002) 

\bibitem{pythia}
T.~Sj\"ostrand, P.~Ed$\acute{\mathrm{e}}$n, C.~Friberg, L.~L\"onnblad, G.~Miu, 
S.~Mrenna and  E.~Norrbin, Comp. Phys. Commun. {\bf 135} (2001) 238 


\end{thebibliography}
\end{document}